\documentclass[lettersize,journal]{IEEEtran}
\usepackage{amsmath,amsfonts}
\usepackage{algorithmic}
\usepackage{algorithm}
\usepackage{array}
\usepackage[caption=false,font=normalsize,labelfont=sf,textfont=sf]{subfig}
\usepackage{textcomp}
\usepackage{stfloats}
\usepackage{url}
\usepackage{verbatim}
\usepackage{graphicx}
\usepackage{cite}
\usepackage{makecell}
\usepackage{multirow}
\usepackage{booktabs}
\usepackage{xcolor}
\usepackage{hyperref}
\usepackage{tikz}
\usepackage{lipsum}
\usepackage[normalem]{ulem}
\begin{document}

\title{Subjective and Objective Quality Assessment of Rendered Human Avatar Videos in Virtual Reality}

\author{Yu-Chih~Chen,
        Avinab~Saha,
        Alexandre~Chapiro,
        Christian~Häne,
        Jean-Charles~Bazin,
        Bo~Qiu,
        Stefano~Zanetti,
        Ioannis~Katsavounidis,        and~Alan~C.~Bovik,~\IEEEmembership{Fellow,~IEEE}
\thanks{Yu-Chih~Chen, Avinab~Saha and Alan~C.~Bovik are with the Laboratory for Image and Video Engineering (LIVE), Department of Electrical and Computer Engineering, The University of Texas at Austin, Austin, TX, 78712 USA (email: berriechen@utexas.edu, avinab.saha@utexas.edu, bovik@ece.utexas.edu).}
\thanks{Alexandre~Chapiro, Christian~Häne, Jean-Charles~Bazin, Bo~Qiu, Stefano~Zanetti, Ioannis~Katsavounidis, are with Meta Platforms Inc., 1 Hacker Way, Menlo Park, CA 94025 USA (email: achapiro@meta.com, chaene@meta.com, jcbazin@meta.com, qiub@meta.com, stefanoz@meta.com, ikatsavounidis@meta.com).}
\thanks{The work of Alan C. Bovik was supported by the National Science Foundation AI Institute for Foundations of Machine Learning (IFML) under Grant 2019844. This work was supported by Meta Platforms Inc.}
\thanks{All experiments, data collection, and processing activities were conducted by the University of Texas at Austin.}}

\newcommand\copyrighttext{%
  \footnotesize \textcopyright © 2024 IEEE. Personal use of this material is permitted. Permission from IEEE must be obtained for all other uses, in any current or future media, including reprinting/republishing this material for advertising or promotional purposes, creating new collective works, for resale or redistribution to servers or lists, or reuse of any copyrighted component of this work in other works.
  DOI: \href{https://doi.org/10.1109/TIP.2024.3468881}{10.1109/TIP.2024.3468881}}
\newcommand\copyrightnotice{%
\begin{tikzpicture}[remember picture,overlay]
\node[anchor=south,yshift=10pt] at (current page.south) {\fbox{\parbox{\dimexpr\textwidth-\fboxsep-\fboxrule\relax}{\copyrighttext}}};
\end{tikzpicture}%
}

\maketitle
\copyrightnotice
\begin{abstract}

We study the visual quality judgments of human subjects on digital human avatars (sometimes referred to as ``holograms" in the parlance of virtual reality [VR] and augmented reality [AR] systems) that have been subjected to distortions. We also study the ability of video quality models to predict human judgments. As streaming human avatar videos in VR or AR become increasingly common, the need for more advanced human avatar video compression protocols will be required to address the tradeoffs between faithfully transmitting high-quality visual representations while adjusting to changeable bandwidth scenarios. During transmission over the internet, the perceived quality of compressed human avatar videos can be severely impaired by visual artifacts. To optimize trade-offs between perceptual quality and data volume in practical workflows, video quality assessment (VQA) models are essential tools. However, there are very few VQA algorithms developed specifically to analyze human body avatar videos, due, at least in part, to the dearth of appropriate and comprehensive datasets of adequate size. Towards filling this gap, we introduce the LIVE-Meta Rendered Human Avatar VQA Database, which contains 720 human avatar videos processed using 20 different combinations of encoding parameters, labeled by corresponding human perceptual quality judgments that were collected in six degrees of freedom VR headsets. To demonstrate the usefulness of this new and unique video resource, we use it to study and compare the performances of a variety of state-of-the-art Full Reference and No Reference video quality prediction models, including a new model called HoloQA. As a service to the research community, we publicly releases the metadata of the new database at \href{https://live.ece.utexas.edu/research/LIVE-Meta-rendered-human-avatar/index.html}{https://live.ece.utexas.edu/research/LIVE-Meta-rendered-human-avatar/index.html}.

\end{abstract}


\section{Introduction}
\IEEEPARstart{R}{ecent} advancements in head-mounted displays (HMDs) and extended reality (XR) technologies have made it possible for people to engage in impressive immersive 3D experiences. Among these, virtual reality (VR) applications now allow for virtual work meetings \cite{mcveigh2022beyond}, entertainment, gaming, and education. This ``metaverse'' of virtual possibilities is also envisioned to include rich augmented reality (AR) applications, whereby users can visualize and interact with the real world, supplemented by graphical overlays, inserted objects, realistic 3D avatars or digital twins, and the ability to communicate with one another by sound and sight from afar \cite{10.3389/frsip.2023.1193523, min2024perceptualvideoqualityassessment}. A major goal in this direction is to give users the ability to interact more intimately and personally via high-quality wireless communications-enabled shared VR or AR experiences.

In one promising scenario relevant to both workplace and social scenarios, two or more users can visually and remotely interact in a virtual space, each person represented by an animated avatar, or ``hologram,'' having facial expressions, body movements, and hand gestures that replicate those of the actual participants. Indeed, as computing hardware, image capture, and graphical processing technologies have advanced, methods for creating high quality, realistic human avatars have significantly improved \cite{li20193d, bagautdinov2021driving}. These 3D human avatar videos are normally represented as dynamic 3D meshes with textured color surfaces. Human avatar model creation involves capturing multiple images from different viewpoints of real world objects or people using high-speed cameras, then reconstructing them into mesh or point cloud geometric and color representations that can be rapidly rendered for 3D visualization. Here we focus on textured mesh data, although point clouds are easily converted into 3D polygonal mesh formats. As 3D digital twins or human avatars become increasingly realistic and data heavy, and as HMDs continue to have improved in space-time resolutions, visual data streams having much higher bandwidths will require perceptually optimized compression tools to ensure efficient, high quality throughput and visualizations.

The possibility of practical human avatar data transmission over the wireless internet presents significant practical challenges. To be able to communicate high quality 3D human avatar videos in real time, volumetric compression is required, which can adversely impact the perceived visual quality of immersive 3D experiences. Since real-time communication is required to enable seamless human interactions, the compression and transmission protocols must minimize temporal and rotational latencies to minimize sensations of lag. Indeed, proposed metaverse infrastructures typically aim to constrain the end-to-end delay (from cloud to client) to less than 20 ms \cite{sun2019communications}.

Further, visual artifacts arising from compression and rendering can significantly degrade the visual quality of human avatar videos. Motivated by the successes of streaming and social media platforms that perceptually optimize video throughput using video quality prediction models, it is highly desirable to be able to quantitatively model human avatar video quality. Given models, video quality prediction algorithms may be devised to measure and control tradeoffs between perceptual human avatar quality and data bandwidth, by providing feedback that can be used to adjust encoder settings.

Assessing the quality of human avatar videos presents unique challenges as compared to traditional video formats. Traditional 2D video quality assessment (VQA) methods are insufficient for human avatar videos, which require accounting for the added complexities of depth perception, spatial resolution, and the dynamic nature of 3D meshes. The subjective experience of users in VR is influenced not only by the visual fidelity but also by the interactivity and realism of the human avatars, necessitating new approaches to VQA that can capture these multifaceted aspects.

VQA research encompasses two general categories of study: subjective and objective video quality. Subjective studies involve the collection of substantial amounts of human subjective judgments of video quality \cite{zhai2020perceptual}. Subjective quality datasets can be used to create, compare, and benchmark objective video quality models. However, existing databases of volumetric picture data are quite limited in terms of their sizes, content varieties, and varieties of realistic distortions. Many of these have deployed 2D displays when conducting human subjective studies, thereby restricting viewpoint, depth sensation, and immersion. Few studies have employed HMDs which allow the study participants to experience immersion in VR or AR \cite{8122237, 8551498, 10.1145/3343036.3352493, gutierrez2020quality, 9089539, 9123121, 9502695, 9252120}.

Conducting studies of VR or AR using modern headsets is important, because the effects of experiencing immersion, six degrees of freedom (6DoF) of movement, and wide-field video compositions can be accounted for. The developments of modern VQA models suitable for analyzing and predicting the quality of VR and AR videos, including ``human avatar'' videos, requires the creation of subjective quality databases that capture these aspects.

Given the urgent need to improve the quality of immersive experiences in VR and AR, and to support the development of robust VQA models, we have created a new psychometric resource called the LIVE-Meta Rendered Human Avatar VQA database. This database addresses the limitations of existing datasets by providing 720 videos derived from 36 source sequences of dynamic human human avatar videos, rendered with varying degrees of spatial and temporal distortions, which were viewed and quality rated by 78 human subjects in an immersive 6DoF VR environment, making it a valuable resource for advancing human avatar video streaming. To demonstrate the value of the new subjective dataset, we also evaluated the performances of a variety of state-of-the-art (SOTA) VQA models on it. We also include new human avatar video quality predictors - HoloQA \cite{avinab2023hologramQA} of our own design, and test and compare them on the new dataset. HoloQA leverages recent advances in visual neuroscience, information theory, and self-supervised deep learning to predict the quality of rendered Digital Human Holograms in VR and AR applications. By adopting a Mixture-of-Experts approach, HoloQA captures both low-level pixel quality and high-level content-aware features specific to the human body. This method achieves SOTA performance on the LIVE-Meta Rendered Human Avatar VQA database and demonstrates competitive performance across other digital human hologram databases. The code for HoloQA will be available post peer review.

We summarize our contributions as follows:
\begin{enumerate}
\item \textbf{Largest Most Comprehensive Human Avatar Perceptual Quality Database}: The new LIVE-Meta Human Avatar VQA Database contains 720 distorted and pristine stimuli from 36 different source human avatar videos, all rated by 78 human subjects, making it the largest 3D graphics VQA database.
\item \textbf{Advancing Immersive Human Avatar Video Streaming}: We further advanced progress on understanding the perception and predictability of streamed avtar videos by evaluating a wide variety of SOTA VQA models on it, by comparing their abilities to predict human quality judgments.
\item \textbf{Resource for Developing and Evaluating VQA Algorithms}: The new LIVE-Meta Rendered Human Avatar VQA Database offers a significant and needed resource for developing and evaluating both FR and NR VQA algorithms tailored for human avatar VR videos. We showed that the new database is also useful for analyzing, benchmarking, and designing FR and NR VQA algorithms.
\end{enumerate}

The remainder of the paper is organized as follows. Section \ref{related} provides an overview of previous subjective and objective VQA quality studies on 3D point clouds and meshes. Sections \ref{database} and \ref{study} introduce the processes of content creation for the new human avatar VQA database, and explain the study design protocol and subjective data acquisition processes, respectively. Section \ref{evaluation} shows the value of the new psychometric resource by comparing the performance of a variety of SOTA VQA models on it. We also describe and analyze new models. Finally, Section \ref{conclusion} concludes the paper and discusses potential directions for future work.

\begin{table*}[!t]
\caption{A summary of existing publicly available 3D graphics IQA/VQA databases \textbf{including human figures} with subjective scores obtained using monitors\label{tab:pchuman}}
\centering
\LARGE
\resizebox{\textwidth}{!}{%
\begin{tabular}{|c|c|c|c|c|c|c|c|c|c|c|c|}
\hline
Dataset name & \# Stimulus & \makecell[c]{\#Source contents} & \makecell[c]{\# Ratings\\per Video} & Resolution & Model & Degradation type & Duration & Display Device & Interaction method & Rendering mode\\
\hline
Alexiou et al., 2017 \cite{10.1117/12.2275142} & 99 & \makecell[c]{9 (4 objects\\ + 5 humans)} & 20 & 147K - 14M points & Static & V-PCC, G-PCC & N/A & 27" monitor & Interactive & Point\\
\hline
Torlig et al., 2018 \cite{torlig2018novel} & 63 & \makecell[c]{7 (4 objects\\ + 3 humans)} & 20 & 482K - 857K points & Static & Octree-based compression + JPEG & N/A & 27" Monitor & Interactive & Point\\
\hline
vsenseVVDB, 2019 \cite{zerman2019subjective} & 32 & 2 humans & 19 & 62K-495K points & Dynamic & V-PCC, downsampling & 6.6 sec & 2D monitor & Interactive & Point \\
\hline
M-PCCD, 2019 \cite{8743258} & 244 & \makecell[c]{8 (7 objects\\ + 1 humans)} & 10+7+15 & 150K - 73.8M points & Dynamic & Octree pruning, 3DTK compression & 24 sec & 49", 55" monitors & Passive & Point \\
\hline
vsenseVVDB2, 2020 \cite{9123137} & 136 & 8 humans & 23 & \makecell[c]{VVs: 402K - 406K points\\8i: 729K - 1.06M points} & Static \& Dynamic & \makecell[c]{Mesh: Draco+JPEG\\Point Clouds: G-PCC, V-PCC}  & 10 sec & 24" monitor & Passive & Point, Mesh\\
\hline
Cao et al., 2020 \cite{9200318} & 120 & 4 humans & 22 & 2048$\times$2048 (texture) & Dynamic & \makecell[c]{Mesh: TFAN+FFmpeg+distance\\Point Clouds: V-PCC+FFmpeg+distance} & 10 sec & 24" monitor & Passive & Point, Mesh\\
\hline
Perry et al., 2020 \cite{9191308} & 90 & 6 humans & \makecell[c]{16+15+\\15+27} & 1M points & Static & G-PCC, V-PCC & 12 sec &\makecell[c]{31", 49",\\55" monitors} & Passive & Point\\
\hline
IRPC, 2021 \cite{9257015} & 54 & \makecell[c]{6 (4 objects\\ + 2 humans)} & 18-20 & 272K - 4.8M points & Static & PCL, G-PCC, V-PCC & 10 sec & 23" monitor & Passive & Point\\
\hline
SJTU-PCQA, 2021 \cite{9238424} & 420 & \makecell[c]{10 (4 objects\\ + 6 humans)} & 16 & N/A & Static & \makecell[c]{Octree-based compression, downsampling,\\ color noise, geometry noise} & 15 sec & 21.5" monitor & Interactive & Point \\
\hline
LS-PCQA, 2022 \cite{10.1145/3550274} & 1240 & \makecell[c]{104 (76 objects\\ + 28 humans)} & 16 &  N/A & Static \& Dynamic & \makecell[c]{Color noise, geometry noise, V-PCC, G-PCC,\\ AVS, Octree-based compression} & 20 sec & 21.5" monitor & Interactive & Point\\
\hline
SJTU-H3D, 2023 \cite{zhang2023advancing} & 1120 & 40 humans & 40 & 2048$\times$2048 (texture) & Static & \makecell[c]{Position, UV map, and texture compression,  geometry/color noise,\\ face simplification, and texture downsampling} & 8 sec & 4K iMac monitor & Passive & Mesh\\
\hline
DHHQA, 2023 \cite{10095347} & 1540 & 55 humans (heads) & 20 & 4096$\times$4096 (texture) & Static & \makecell[c]{Surface simplification, position compression, UV compression,\\texture sub-sampling, texture compression, color noise, geometry noise} & N/A & 4K iMac monitor & Passive & Mesh\\
\hline
DDH-QA, 2023 \cite{10219874} & 800 & 2 humans& 41 & 2048$\times$2048 (texture) & Dynamic & \makecell[c]{Color noise, geometry noise, texture compression, texture downsampling, \\position compression, UV map compression, \\skeleton binding error, motion range unnaturalness} & N/A & 4K iMac monitor & Passive & Mesh\\
\hline
\end{tabular}%
}
\end{table*}

\begin{table*}[!t]
\caption{A summary of existing publicly available 3D graphics IQA/VQA databases rated with subjective scores obtained using HMDs\label{tab:pcvr}}
\centering
\Large
\resizebox{\textwidth}{!}{%
\begin{tabular}{|c|c|c|c|c|c|c|c|c|c|c|c|}
\hline
Dataset name & \# Stimulus & \makecell[c]{\#Source contents} & \makecell[c]{\# Ratings\\per Video} & Resolution & Model & Distortion type & Duration & Display Device & Interaction method & Rendering mode\\
\hline
Alexious et al., 2017 \cite{8122237, 8551498} & 40 & 5 objects & 21 & 22K - 36K points & Static & Gaussian noise, octree-pruning & N/A & \makecell[c]{AR\\ (Occipital Bridge)} & Interactive & Point \\
\hline
Nehm\'{e} et al., 2019 \cite{10.1145/3343036.3352493} & 80 & 5 objects & 30 & 250K - 600K points & Static & \makecell[c]{Geometric quantization, color quantization,\\ color distortions} & 6, 10 sec & \makecell[c]{VR\\ (HTC Vive Pro)} & Passive & Mesh\\
\hline
Guti{\'e}rrez et al., 2020 \cite{gutierrez2020quality} & 28 & 4 objects & 24 & N/A & Static & Geometry quantization, JPEG compression & 15 sec & \makecell[c]{MR/AR\\ (Microsoft HoloLens)} & Interactive & Mesh\\
\hline
Subramanyam et al., 2020 \cite{9089539} & 72 & 8 humans & 27+25 & N/A & Dynamic & the MPEG anchor, V-PCC & 5 sec & \makecell[c]{VR\\ (Oculus Rift)} & Interactive & Point, Mesh \\
\hline
PointXR, 2020 \cite{9123121} & 40 & 5 objects & 20 & 4096$\times$4096 & Static & G-PCC & 13.7, 23 sec & \makecell[c]{VR\\ (HTC Vive Pro)} & Interactive  & Point\\
\hline
SIAT-PCQD, 2021 \cite{9502695}& 340 & \makecell[c]{20 (10 objects\\ + 10 humans)} & 38 & 145K - 1.6M points & Static & V-PCC & ~20 sec & \makecell[c]{VR\\ (HTC Vive)} & Interactive & Point\\
\hline
Nehm\'{e} et al., 2021 \cite{9252120} & 480 & 5 objects & 24 & 216K - 1.3M points & Static & \makecell[c]{Geometric quantization, color quantization,\\ color distortions} & 10 sec & \makecell[c]{VR\\ (HTC Vive Pro)} & Interactive & Mesh \\
\hline
\makecell[c]{LIVE-Meta Rendered Human Avatar\\VQA Database, 2023} & 720 & 36 humans & 26 & 2048$\times$2048 (texture) & Dynamic & \makecell[c]{Temporal artifacts, reduced texture resolution,\\ reduced frame rate} & 15 sec & \makecell[c]{VR\\ (Oculus Quest Pro)} & Interactive & Mesh \\
\hline
\end{tabular}%
}
\end{table*}
\section{Related Works} \label{related}

This section reviews previous studies related to the subjective and objective quality assessment of 3D graphics and human avatar videos. We discuss various prior methodologies and their relevance to our research.

\subsection{Subjective 3D Graphics Quality Assessment}

Since 2014, several datasets have been developed and utilized for evaluating the quality of 3D graphics, represented as 3D meshes and point clouds, with and without appearance attributes. A number of research groups have conducted subjective quality assessment tests involving these types of 3D data. Early studies focused on evaluating static object contents displayed on 2D screens \cite{7009910, 8026263, 10.1117/12.2275142, 8463406, torlig2018novel}. Some users of these datasets \cite{8463406, 8026263} converted the original point clouds into polygonal meshes via surface reconstruction methods prior to using them, or vice-versa.

Previous studies primarily concentrated on colorless point clouds and only explored human responses to a limited range of degradations, such as downsampling and noise generation. The SJTU-PCQA database\cite{9238424} introduced additional relevant compression/distortion types, including geometric distortions, Gaussian noise, and octree-pruning. Later, Geo-Metric \cite{10.1145/3550454.3555475} introduced more geometric distortions, including 4 types of noise, smoothing, and simplification.

The emergence of a real-time point cloud codec for 3D immersive video \cite{7434610} in 2017 helped drive research into the development of point cloud quality assessement (PCQA) methodologies. This codec found applications in immersive and augmented communication scenarios, and was later considered as a standardized point-cloud compression solution by MPEG. The Video-based PCC (V-PCC) quality prediction model, which targeted dynamic point clouds, was applicable to colored and rendered point clouds \cite{zerman2019subjective}, leading to its use in evaluating advanced point cloud codecs in subjective PCQA studies. These studies revealed that texture distortions generally had a greater impact on perceived quality than geometric distortions, particularly when evaluating images of human figures.

Later, many databases incorporated dynamic 3D graphics to represent the movement of 3D objects, resulting in temporal content variations and temporal artifacts \cite{zerman2019subjective, 8743258, 9123137, 9200318}, and including moving human figures \cite{zerman2019subjective, 8743258, 9123137, 9238424, 10.1145/3550274, 10219874, 9089539}. 

Subjective comparisons between point clouds and their corresponding reconstructed meshes were first studied in \cite{8463406}, but no definitive conclusions were drawn regarding the superiority of either representation. A later study \cite{9123137} was the first attempt to compare textured meshes and colored point clouds in the context of compression.  This study found that textured meshes tend to exhibit superior quality at higher bitrates, whereas colored point clouds demonstrate enhanced performance in scenarios involving limited bandwidth and storage capacity. Another study \cite{9200318} investigated the combined impacts of viewing distance and bitrate on the perceptual quality of compressed 3D human figure sequences. Their findings suggested that viewers preferred meshes at viewing distances of 1.5 meters, while point clouds were generally favored at lower bitrates. 

Another subjective database, named IRPC \cite{9257015}, was published to investigate the effects of coding and rendering on the perceptual quality of point clouds, without considering color attributes. These datasets were limited in terms of content, distortion types, and representation of prevailing codecs, making them inadequate for developing learning-based PCQA algorithms.

Different subjective evaluation methods, including the Absolute Category Rating with Hidden Reference (ACR-HR), and Double Stimulus Impairment Scale (DSIS) were considered and compared for subjective VR studies \cite{10.1145/3343036.3352493, 9252120}. Their results suggested that DSIS led to better accuracy than ACR-HR, while DSIS participants required more time to rate the videos. 

One novel quality evaluation methodology proposed by Torlig \textit{et al.} \cite{torlig2018novel} allowed the human subjects to interact with the viewed content by zooming, rotating, and translating, using a mouse. Some subsequent studies continued to have subjects passively view and assess perceptual quality \cite{8743258, 9123137, 9200318, 9191308, 9257015, zhang2023advancing, 10095347, 10219874}, other studies allowed subjects to manipulate 3D graphical content on 2D monitors \cite{zerman2019subjective, 10.1117/12.2275142, 9238424, 10.1145/3550274}. However, these interactive datasets usually only encompassed static objects.

Towards providing more realistically immersive user experiences in human studies, several researchers employed 3D visualization tools, including HMDs, with subjects operating in VR and AR environments \cite{10.1145/3343036.3352493, 9089539, 9123121, 9502695, 9252120, 8122237, 8551498, gutierrez2020quality}, including 3DoF and 6DoF VR environments \cite{9502695}. Subramanyam \textit{et al.} \cite{9089539} compared viewing conditions enabling 3DoF and 6DoF VR and developed the PointXR toolbox for conducting PCQA in VR environments \cite{9123121}. These studies explored different aspects of subject mobility, fixed positioning, navigation using head movements alone with 3DoF \cite{9089539}, and free body navigation in a room with 6DoF \cite{9123121}. One study focused on subjective PCQA in 6DoF VR environments \cite{9502695}, resulting in a subjective database called SIAT-PCQD \cite{9502695} with compression-induced combinations of geometric and texture distortions. They also proposed two projection-based objective quality evaluation methods.

To our knowledge, just seven previous 3D graphical human studies have been conducted using VR and AR HMDs \cite{8122237, 8551498, 10.1145/3343036.3352493, gutierrez2020quality, 9089539, 9123121, 9502695, 9252120}, and only one of them includes dynamic human figures \cite{9089539}, as indicated in Table \ref{tab:pcvr}. There are no 3D graphical subjective quality datasets that include rendered human avatars impaired by spatial and temporal distortions, viewed by human subjects in a 6DoF VR environment. Currently available datasets are insufficient to conduct studies of such deep, immersive experiences. Therefore, there is a need for such datasets, which are the kind of important scientific tools that are needed to develop AR/VR quality prediction models, which in turn are needed to perceptually optimize processing protocols such as rendering, scaling, and compression. Towards addressing these needs, we have created such a perceptual resource, which we call the LIVE-Meta Rendered Human Avatar VQA Database. We describe the details of construction, content, and experimental design of this new dataset in the following sections. However, Table \ref{tab:pchuman} supplies a basic comparison of existing 3D graphics datasets in terms of size, content, distortion types, and display devices.

\subsection{Objective 3D Graphics Quality Assessment}

Over the last decade, numerous PCQA and mesh compression datasets have been developed, leading to the introduction of several objective quality assessment models specifically designed for 3D graphics. Next we discuss objective video quality assessment models designed specifically for the analysis of point cloud and mesh videos.

\subsubsection{Objective Point Cloud Quality Assessment Models}\label{sec:object-metric}
FR VQA models are commonly used to evaluate the quality of point clouds. These FR models can in turn be classified as point-based models or as projection-based models \cite{torlig2018novel}. They can also be further categorized by the type of distortion type being evaluated, whether geometric texture-based, or some combination.

Point-based FR VQA models have been proposed to evaluate specific types of point cloud distortions, such as geometry and color. While these models offer the advantage of computing explicit information that can be stored in point cloud formats and have been utilized in recent studies, they have been found to poorly predict visual quality across different types of content \cite{zhang2022mm}.

Projection-based models project both reference and test point clouds onto six planes. This allows the application of conventional 2D objective video quality models to directly measure geometric and color artifacts \cite{torlig2018novel}. The processes of real-time voxelization and projection also reduce the computational complexity of VQA.

For example, in \cite{9238424}, a projection-based approach was employed where 3D point clouds were projected onto six perpendicular cubic faces. Weights were then computed and applied on color texture and depth images from the different projection planes, then summed to generate the final quality index. Excluding pixels that belong to the background can improve the accuracy of quality prediction \cite{8743277}. Increasing the number of projected views only moderately improves predictions, while incorporating user interactivity information can enhance performance \cite{8743277}. When viewing inanimate objects, viewers take longer to access the content; when viewing human body models, frontal and face views consistently receive more attention \cite{10.1145/1877808.1877819}.

Consequently, frontal views are often deemed the optimal configuration in human body datasets, while a greater variety of perspectives better represents scenes containing inanimate objects. However, models based on these premises tend to be viewpoint-dependent \cite{7272102}.


In point cloud applications, it is often impractical to obtain original point clouds because of storage limitations and inadequate communication bandwidths. In such instances, using FR VQA models may be infeasible. NR quality assessment models, which estimate point cloud quality without the availability of original point clouds, are the necessary instruments in such applications. 



When evaluating the visual quality of point clouds, it is important to consider both static and dynamic aspects, as humans visualize point clouds simultaneously in space and time. Consequently, VQA models that can account spatio-temporal content/distortions offer the greatest potential quality prediction power. One such model called VQA-PC \cite{zhang2022treating} utilizes a trainable 2D-CNN and pre-trained 3D-CNN modules to extract spatial and temporal features. By treating point clouds as moving camera videos, VQA-PC advanced the field of projection-based PCQA, leveraging both static and dynamic views.




\subsubsection{Objective Mesh Quality Assessment Models}

Algorithms that have been designed to evaluate the visual quality of 3D meshes can be generally categorized into two types: model-based ones, which operate directly on the 3D models \cite{7283644, wang2012fast, lavoue2011multiscale}, and image quality assessment (IQA) models which operate on rendered snapshots of 3D models \cite{yang2004optimized, caillaud2016progressive}. Most model-based approaches lead to FR models, and are inspired by successful IQA models. Model-based NR mesh quality algorithms \cite{abouelaziz2016no, 7907557, 8296382} are generally able to predict the quality of 3D meshes when there are geometric distortions, but usually do not address color distortions. However, the NR model in \cite{9455963} operates on 3D color meshes by extracting quality-aware geometric and color features which are integrated into quality scores using a support vector regressor (SVR). This model was extended to assess both 3D colored point cloud and mesh models in \cite{9949359}, whereby quality-aware features are extracted using 3D natural scene statistics and entropy of geometric and color information. These features are then processed using an SVR to predict perceived quality, providing a robust framework for NR quality assessment of various 3D content types.

A variety of deep learning-based models trained to analyze the quality of 3D meshes have been proposed. Abouelaziz \textit{et al.} \cite{8296382} train a CNN using hand-crafted perceptual geometric features extracted from 3D meshes. They also proposed a later model that extracts feature vectors along 3 different CNN paths, then combines them \cite{abouelaziz2020no}. However, these deep models only consider geometric meshes without analyzing color/texture attributes. The authors of \cite{10.1145/3592786} trained a learning-based NR model for predicting the quality of textured meshes, using a large-scale dataset of more than 343k textured meshes. Their NR models are image-based; hence their approach takes the approach of transferring the complex task of mesh quality prediction to the simpler one of IQA.

\section{LIVE-Meta Rendered Human Avatar VQA Database} \label{database}

\begin{figure}[!t]
	\centering
	\footnotesize
	\renewcommand{\tabcolsep}{1.3pt} 
	\renewcommand{\arraystretch}{1.3} 
	\begin{tabular}{c cc cc}
        \includegraphics[height=90pt]{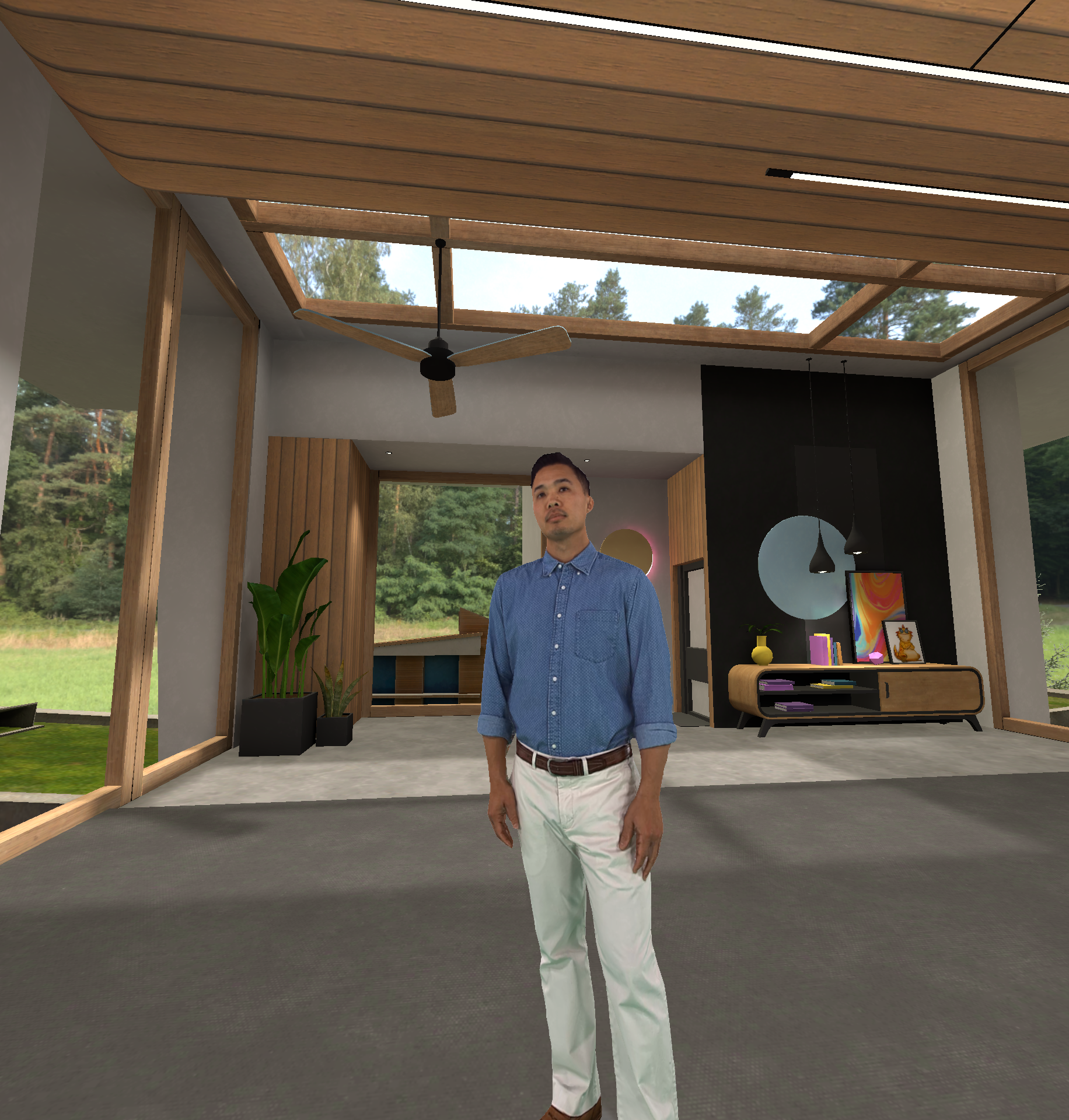} & & & &
        \includegraphics[height=90pt]{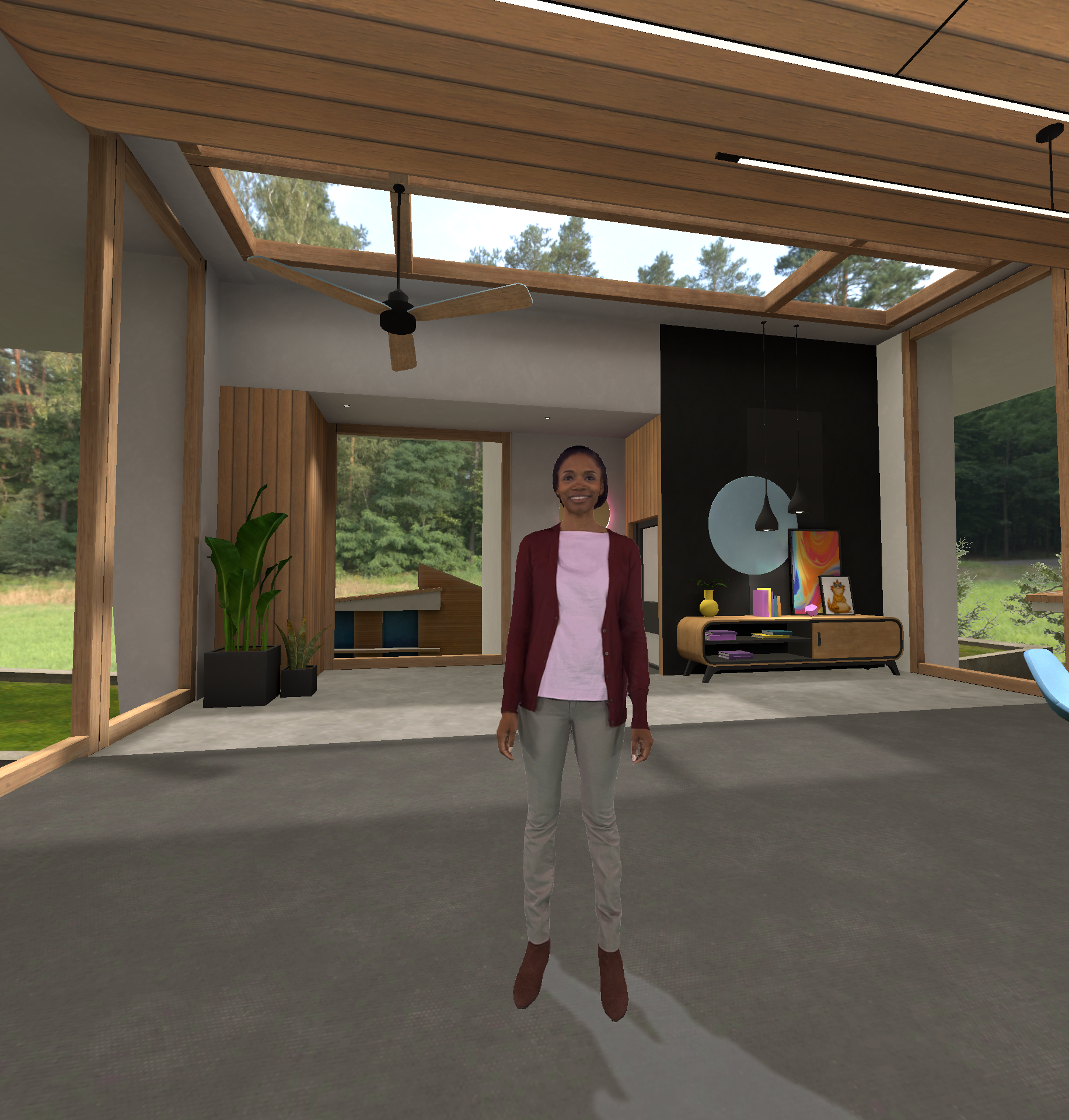}\\
        (a) 27-Carl Standing Business & & & & (b) 09-Robyn Casual Talking\\        
        \includegraphics[height=90pt]{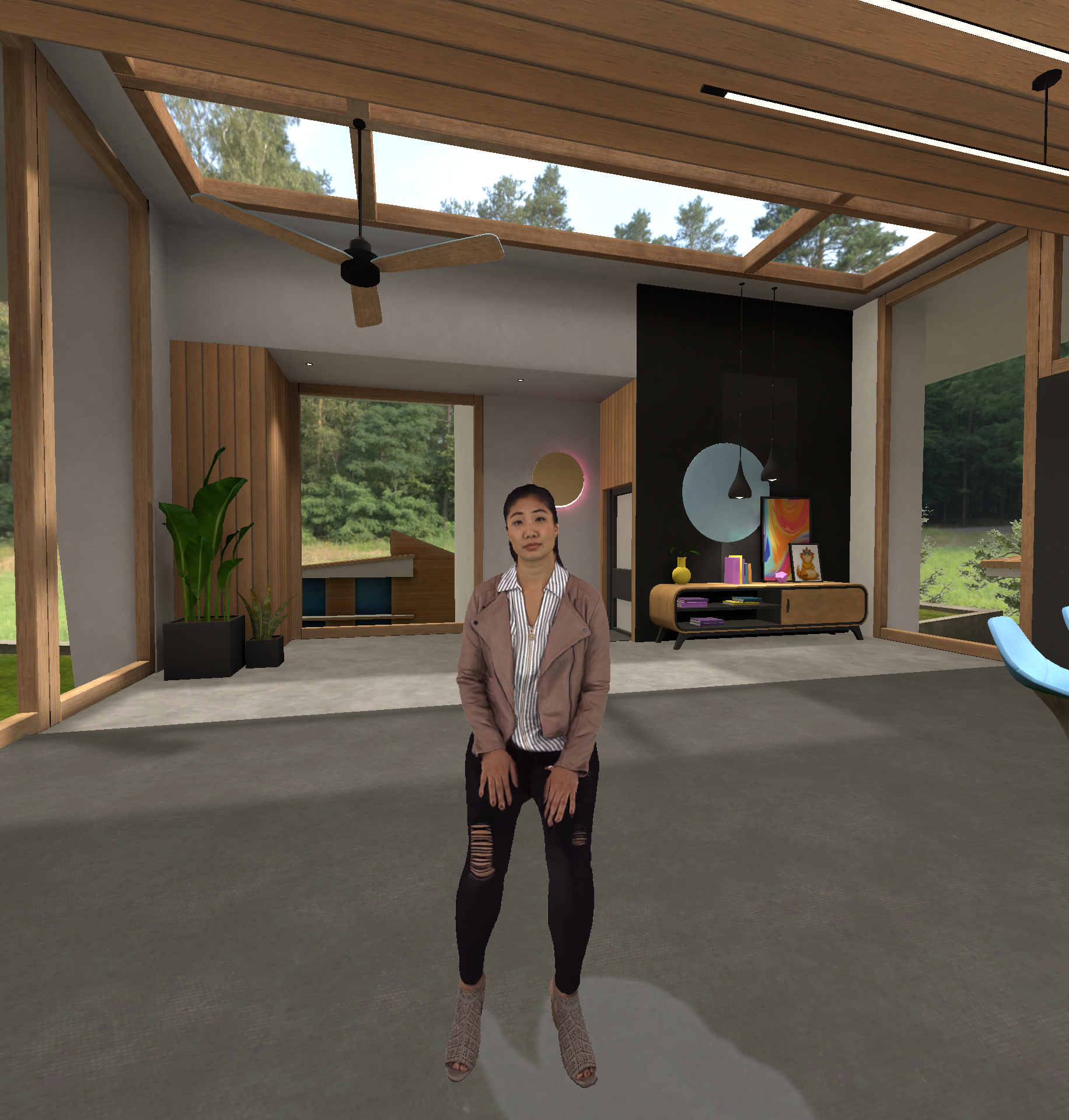} & & & &
        \includegraphics[height=90pt]{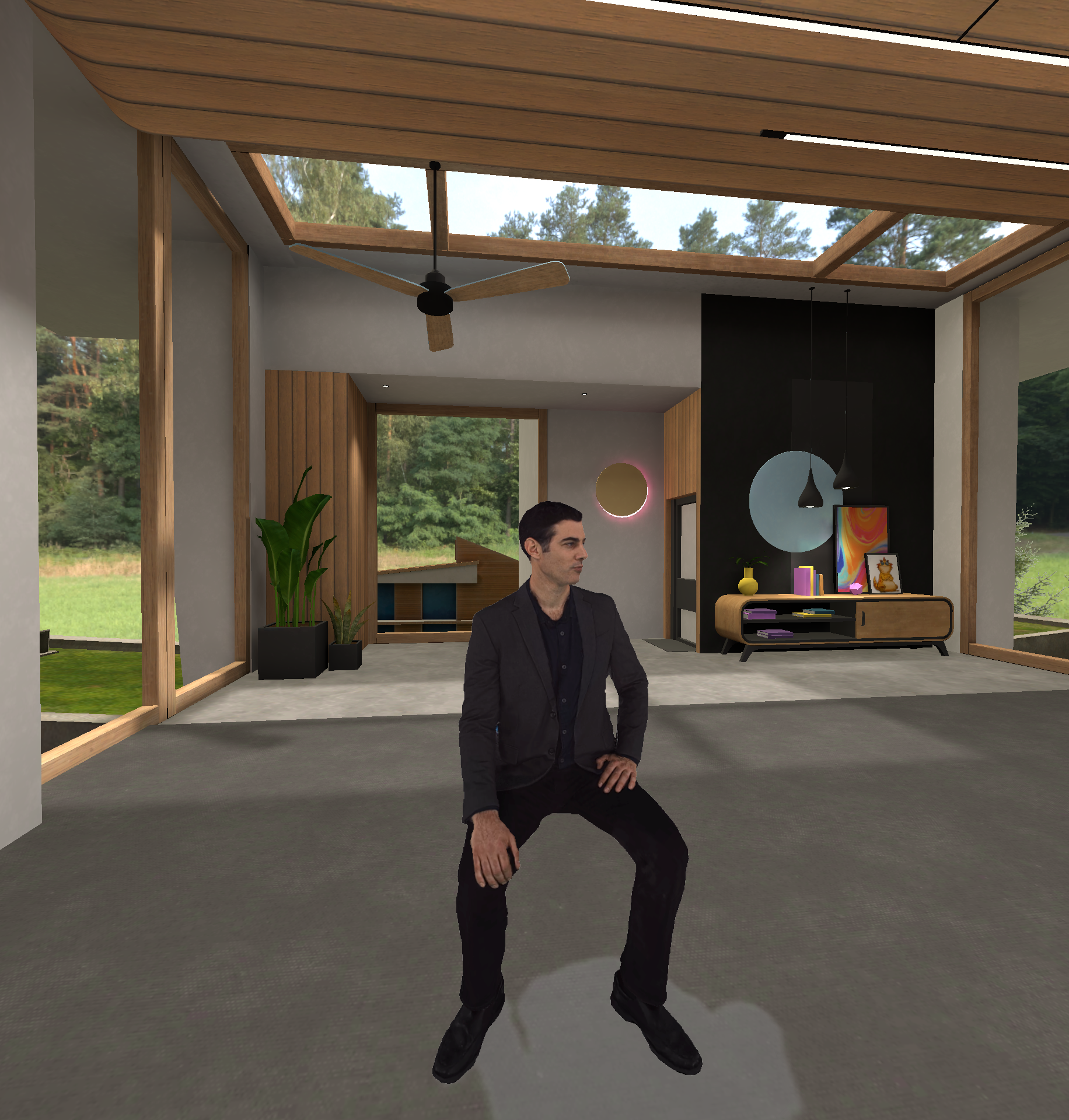}\\
        (c) 21-Julia Seated Talking & & & & (d) 25-Luke Seated Small Talk\\
	\end{tabular}
   \caption{Sample frames of (a) (b) standing and (c) (d) sitting human avatar videos from the LIVE-Meta Rendered Human Avatar VQA Database.}
	\label{fig:pose}
\end{figure}

The LIVE-Meta Rendered Human Avatar VQA Database consists of 720 video sequences created by adding compression artifacts to 36 pristine human avatar videos using 20 different encoding parameter settings. These videos were used as stimuli in a laboratory-based human subjective study of human avatar video quality. Several sample frames depicting standing and sitting poses from the human avatar videos are shown in Fig. \ref{fig:pose}. Next, we provide a detailed description of the dataset preparation, including obtaining the source sequences, protocol for adding artifacts, and the volumetric simulation-rendering pipeline.

\subsection{Source Sequences}
We purchased 36 pristine human avatar videos from the Metastage shop, tabulated by title in Table \ref{tab:content}. Metastage recorded individuals from various angles using 106 cameras and reconstructed their actions and emotions into 3D Unity assets at a texture resolution of 2048$\times$2048 wrapped with MP4 mesh textures. Each human avatar has a mesh polycount of approximately 20,000 triangles.

The original durations of the 36 reference videos, which ranged from 14 to 32 seconds, was clipped to 14 to 15 seconds to facilitate practical use in the human study. Through a trial study involving three participants familiar with VR headsets, and two more participants unfamiliar with them, we determined that 15 second durations are sufficient to enable subjects to rate the video quality. Two of the participants experienced dizziness after approximately 30 minutes of viewing, so we avoided sessions longer than this to prevent feelings of discomfort caused by using the VR headsets.

To ensure a balanced representation, the 36 videos were selected based on demographic characteristics, attire colors, and poses including standing and sitting but excluding walking and dancing. Subjects having distracting accessories, such as doctor's stethoscope, were excluded. As shown in Table \ref{tab:content}, the set of pristine videos were divided into six video groups, each containing a diverse mix of genders, skin colors, poses, and clothing colors. Videos of the same individual were allocated to different groups, to prevent repetition in any of the sessions of the human study, as described in Section \ref{sec:subjects}.

\begin{table*}[!t]
\caption{Groupings of Metastage Shop human avatar Videos with diverse attributes of gender, skin color, poses and clothing, to avoid repetition within sessions}\label{tab:content}
\centering
\normalsize
\resizebox{0.9\textwidth}{!}{%
\begin{tabular}{|c|c|c|c|c|c|}
\hline
\multicolumn{2}{|c|}{Study Group A} & \multicolumn{2}{c|}{Study Group B} & \multicolumn{2}{c|}{Study Group C}\\
\hline
Video Group 1 & Video Group 2 & Video Group 3 & Video Group 4 & Video Group 5 & Video Group 6\\
\hline
\makecell[l]{01-Natasha Serious Talking\\
06-im Listening Casual\\
07-Robyn Seated Talking\\
11-Wendy Listening Casual\\
13-Terence Seated Listening\\
16-Frank Listening Casual} & 
\makecell[l]{04-Jim Listening Business\\
19-Deena Listening Casual\\
22-Julia Serious Talking\\
24-Doctor Luke Seated Listening\\
28-Carl Listening Business\\
32-Jenny Sitting Casual} &
\makecell[l]{02-Natasha Seated Listening\\
08-Robyn Listening Casual\\
14-Frank Casual Talking\\
20-Deena Casual Talking\\
26-Luke Seated Listening Party\\
31-Amanda Seated Listening} &
\makecell[l]{05-Jim Standing Business\\
17-Deena Listening Business\\
23-Luke Study Chart\\
29-Carl Serious Talking\\
34-Jenny Seated Listening\\
36-Sophie Typing
} &
\makecell[l]{03-Jim Serious Talking\\
09-Robyn Casual Talking\\
10-Wendy Listening Business\\
12-Terence Seated Talking\\
15-Frank Standing Casual\\
21-Julia Seated Talking
} &
\makecell[l]{18-Deena Presentation\\
25-Luke Seated Small Talk\\
27-Carl Standing Business\\
30-Amanda Listening\\
33-Jenny Casual Talking\\
35-Sophie Seated Typing
}\\
\hline
\end{tabular}
}
\end{table*}

\subsection{Video Distortions}\label{sec:adding_artifacts}

Table \ref{tab:distortion} tabulates the 20 different distortion settings that are meant to simulate events that might occur during cloud streaming, yielding a total of 720 videos. These were generated using a special-purpose application tool which we will introduce in Section \ref{study}. The Table shows that the distortions are indexed from 1 to 20. This includes the 36 pristine reference videos (index = 1), each of which was processed with eight single distortions (index = 2 to 9) and 11 combinations of multiple distortions (index = 10 to 20).

Among the single distortions, delay artifacts arise when the presumed viewing angle at which the mesh is generated differs from the viewer's actual current viewing angle because of processing or communication latencies. The delays ranged from 100 ms to 300 ms, which are tolerable in human avatar video streaming. Delays exceeding 400 ms could significantly degrade user experiences, leading to potential user disengagement in real-world scenarios. Distortions from color resolution/scaling ranged from 480p to 2048p (pristine), while frame rates were varied over 10 fps to 30 fps. The delay distortions present visually as temporal self-occlusions, resulting in unnatural visual overlaps. Distortions from color resolution/scaling manifest as blur or blockiness. From our observations, a color resolution at 1600p is the threshold where distortions start becoming noticeable. Specifically, 1080p is considered fair quality, 960p is between fair and poor, 720p is poor, 600p is between poor and bad, and 480p is bad, representing the worst-case scenario we considered. Frame rate distortions cause visual sensations of temporal discontinuity or ``statter,'' especially when there is rapid motion. Frame rates of 20 fps or 15 fps are generally tolerable, but at 10 fps, the quality is significantly impacted.

The combinations of multiple distortions are divided into three categories: ``moderately limited bandwidth conditions,'' (MLBC) ``heavily limited bandwidth conditions,'' (HLBC) and ``severely limited bandwidth conditions'' (SLBC), as shown in Table \ref{tab:distortion}. These are combinations of the single distortions, but also include distortions from reduced depth resolution. Our choice of parameters resulted in a diverse and wide range of perceptual distortions and contents to ensure noticeable differences between the various distortion levels. This diversity was intended to cover a broad spectrum of real-world scenarios and to thoroughly evaluate the robustness of video quality assessment models.

Regarding the quality of the pristine reference videos (index = 1), due to limitations of the Metastage capture system, human avatar reference videos were reconstructed to closely resemble the original source human avatar videos. Although the capture device and reconstruction technology impose certain limitations, these reference videos are considered pristine for the purpose of our study. The background used in these videos is a high-quality static PNG image, which does not significantly impact the overall evaluation of the photorealistic foreground avatars. Our statistical post-analysis in Section \ref{sec:subject-consistency-analysis} indicates that this combination was acceptable to the study participants.
\begin{figure*}[ht!]
    \centering
    \includegraphics[width=\textwidth]{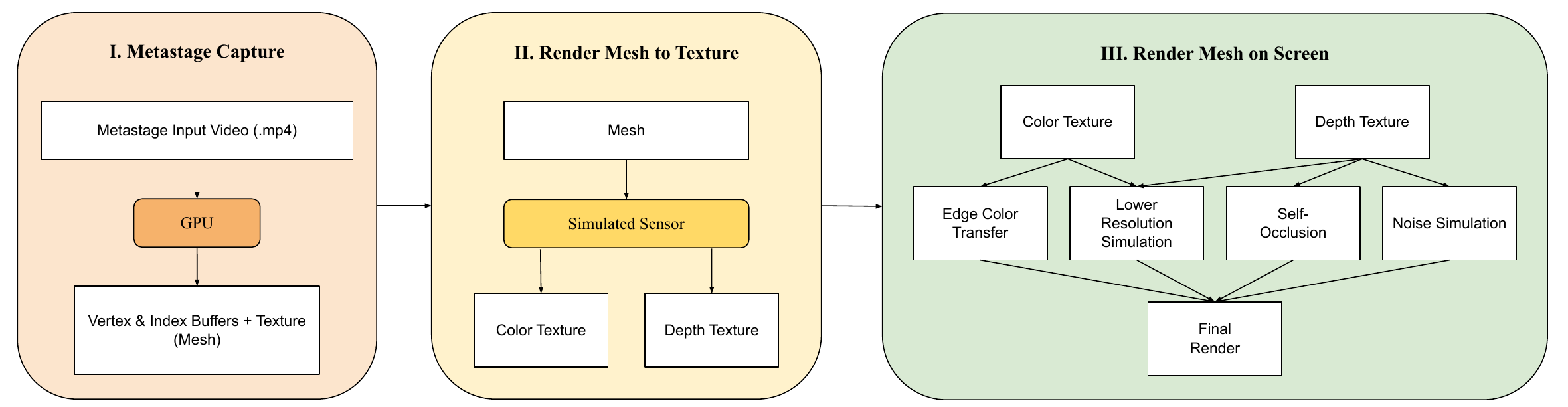}
    \caption{A simplified flow diagram of the volumetric simulation-rendering pipeline used in the privately released Unity tool provided by the Meta platform. This diagram highlights the key stages in the conversion from raw mesh data to RGB-D, including the loading of Metastage mesh, rendering from the sensor's perspective, and reprojecting textures from the headset's point of view.}%
    \label{fig:pipeline}
\end{figure*}
\subsection{Volumetric Simulation-Rendering Pipeline}

The conversion from Metastage raw mesh data to RGB-D was accomplished via a volumetric simulation-rendering pipeline, a simplified model of which is shown in Fig. \ref{fig:pipeline}.

Initially, the Metastage mesh was loaded via the Metastage Unity plugin which we will introduce in Section \ref{sec:tool}. The plugin output was a mesh with texture animating at 30 fps. The mesh integrated into the Unity rendering pipeline, allowing placement in the scene and correct depth rendering relative to other 3D objects. To better anchor the representation in the scene, a shadow effect from the mesh to the floor was applied. Additionally, the mesh representation allowed the OVR plugin, which is a tool that facilitates the integration of virtual reality features into Unity projects, to seamlessly render them stereoscopically in the VR headset, providing users with the experience of viewing a person in front of them.

The input .mp4 files were stored in the streaming assets folder of the Unity project. This allowed users to modify the video files without modifying and building the application. Users of the Unity application can add new .mp4 videos and update the configuration file, and the application will load them dynamically. The video data was processed by the GPU and loaded into vertex and index buffers. This process was internally controlled by the Metastage plugin. The .mp4 format used by Metastage is a proprietary extension that includes vertices, indices, and texture. Once the data was loaded into GPU buffers by the plugin, we could render it. The bitrate was relatively high, as Metastage decoded and read every vertex and triangle per frame of the capture. The reconstructed mesh typically consisted of 15,000 to 30,000 vertices and 20,000 to 40,000 triangles per frame, with a standard frame rate of 30 fps.

Subsequently, the GPU buffers containing the mesh, as filled by Metastage, were rendered using a simple unlit shader. We rendered the mesh from the perspective of the sensor to simulate various artifacts in a later stage. This render pass generated two textures: color and depth. The color texture simulated what the sensor would observe, while the depth texture simulated what an ideal depth sensor would perceive. Both color and depth resolutions could be toggled.

Finally, with the color and depth textures available, we rendered the Metastage mesh again from the headset's point of view. We utilized a custom shader that reprojected the previously produced color and depth textures onto the mesh, taking into account the sensor's perspective. Additionally, noise could be added to the sampled depth data to simulate a physical depth sensor. This data was then used to modify the Metastage mesh to conform to the sampled depth data. The depth texture was also employed to calculate self-occlusion from the sensor's perspective. Instead of using the original high-fidelity Metastage texture, the color texture was sampled to simulate lower resolution and edge color transfer. It is important to note that if we wish to simulate any topical artifacts or treatments, we must provide data to the shader so that it can identify areas of interest in 3D space (UV space is not useful due to Metastage encoding technology jitter).

\begin{table}[!t]
\caption{List of 20 distortion settings used in the study, covering a range of delays, color resolutions, and frame rates.}\label{tab:distortion}
\centering
\normalsize
\resizebox{\columnwidth}{!}{%
\begin{tabular}{|c|c|c|c|c|c|c|}
\hline
Index & Delay (ms) & \makecell[c]{Depth\\Error (m)} & \makecell[c]{Depth\\Resolution} & \makecell[c]{Color\\Resolution} & \makecell[c]{frame rate Mean\\$\pm$Variance (fps)} & Distortion Type\\
\hline
1 & 0 & 0 & 1000p & 2048p & 30 & pristine\\
\hline
2 & 100/200 & 0 & 1000p & 2048p & 30 & delay\\
\hline
3 & 300/400 & 0 & 1000p & 2048p & 30 & delay\\
\hline
4 & 0 & 0 & 1000p & 1600p/1280p & 30 & color resolution/scaling \\
\hline
5 & 0 & 0 & 1000p & 1080p/720p & 30 & color resolution/scaling\\
\hline
6 & 0 & 0 & 1000p & 640p/480p & 30 & color resolution/scaling\\
\hline
7 & 0 & 0 & 1000p & 2048p & 20 & frame rate reduced\\
\hline
8 & 0 & 0 & 1000p & 2048p & 10 & frame rate reduced\\
\hline
9 & 0 & 0 & 1000p & 2048p & random 15$\pm$10 & frame rate reduced\\
\hline
10 & 100 & 0.025 & 480p & 1600p & 30 & MLBC\\
\hline
11 & 200 & 0.025 & 480p & 1920p & 30 & MLBC\\
\hline
12 & 300 & 0.025 & 480p & 1600p & 30 & MLBC\\
\hline
13 & 100 & 0.05 & 360p & 1280p & 30 & HLBC\\
\hline
14 & 100 & 0.05 & 360p & 1080p & 30 & HLBC\\
\hline
15 & 100 & 0.05 & 360p & 1080p & 15 & HLBC\\
\hline
16 & 400 & 0.05 & 360p & 1280p & 30 & HLBC\\
\hline
17 & 400 & 0.05 & 360p & 1080p & 30 & HLBC\\
\hline
18 & 400 & 0.05 & 360p & 1080p & 15 & HLBC\\
\hline
19 & 100 & 0.075 & 120p & 720p & 15 & SLBC\\
\hline
20 & 300 & 0.075 & 120p & 720p & 15 & SLBC\\
\hline
\end{tabular}%
}
\end{table}

\subsection{Dataset and Metadata Description}
The dataset includes the following components:
\begin{enumerate}
\item \textbf{Human Avatar Videos}: A set of 36 high-quality human avatar videos purchased from Metastage, categorized by gender, attire, skin color, movement, and duration. Although we cannot make the proprietary Metastage videos freely available, interested readers may also purchase them.
\item \textbf{Tool Instructions}: Detailed instructions on using the Unity Binary Tool for running the application, displaying video playlists, and visualizing artifacts.
\item \textbf{Asset List}: A comprehensive list of all the Metastage assets used in the study, detailing the characteristics and categorization of each video asset.
\item \textbf{Metadata}: Detailed metadata for each video, including timestamps, frame information, user ratings, and artifact configurations.
\end{enumerate}

The metadata structure includes:
\begin{itemize}
\item \textbf{Video Information}: Details about each video, including filename, duration, and content type.
\item \textbf{Experiment Logs}: Logs of user interactions, ratings, and head movements during the experiments.
\item \textbf{Configuration Files}: JSON files used to configure and control the video playback and artifact simulations.
\end{itemize}
By providing this detailed description, we aim to enhance the transparency and reproducibility of our research.

\section{Human Study Design} \label{study}

This section provides a comprehensive description of the subjective quality assessment study, including details about the test environment, the interface of the assessment tool, the experimental protocol, the evaluation methodology, the post-study questionnaire, and analysis and processing of the subjective scores.

\begin{figure*}[!ht]
	\centering
	\normalsize
	\renewcommand{\tabcolsep}{1pt} 
	\renewcommand{\arraystretch}{1} 
	\captionsetup[subfloat]{labelfont=footnotesize,textfont=footnotesize}
    \subfloat[Initial State of Tool\label{fig:initial-state}]{\includegraphics[width=0.3\textwidth]{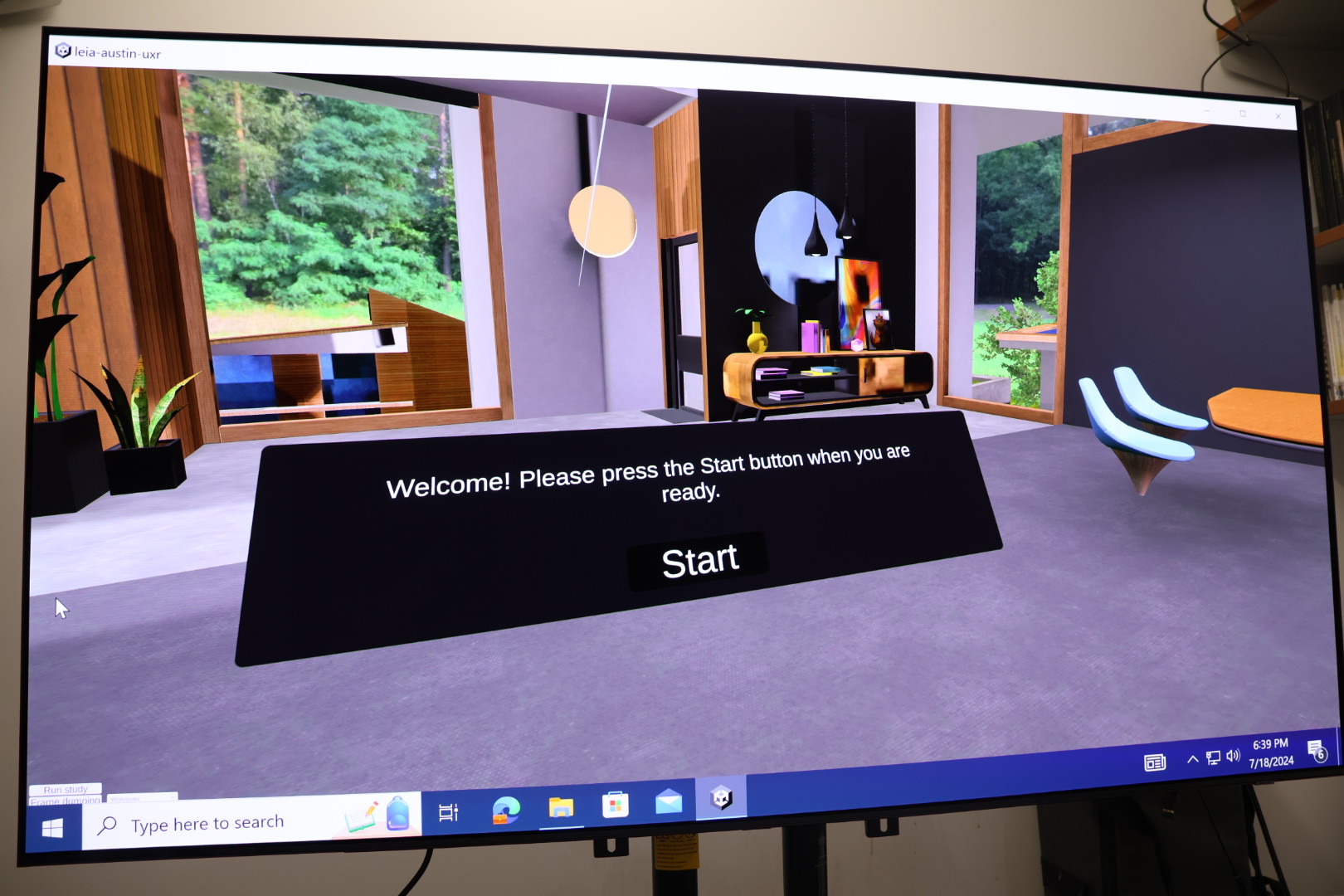}}
    \hspace{10pt}
     \subfloat[Lab Settings and Human Avatar Video Playback\label{fig:video-playback}]
     {\includegraphics[width=0.3\textwidth]{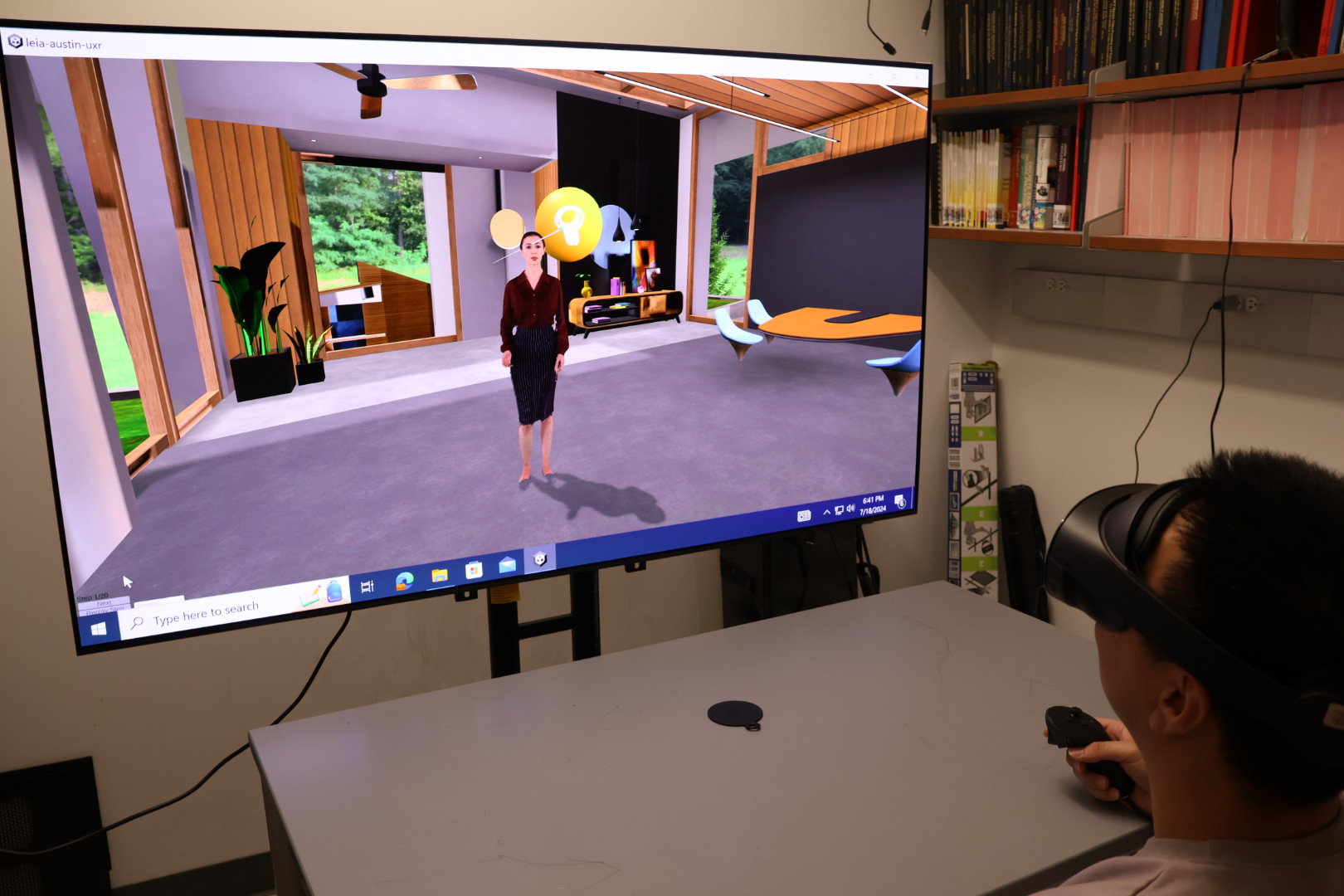}}
     \hspace{10pt}
     \subfloat[Rating Bar\label{fig:rating-bar}]{\includegraphics[width=0.3\textwidth]{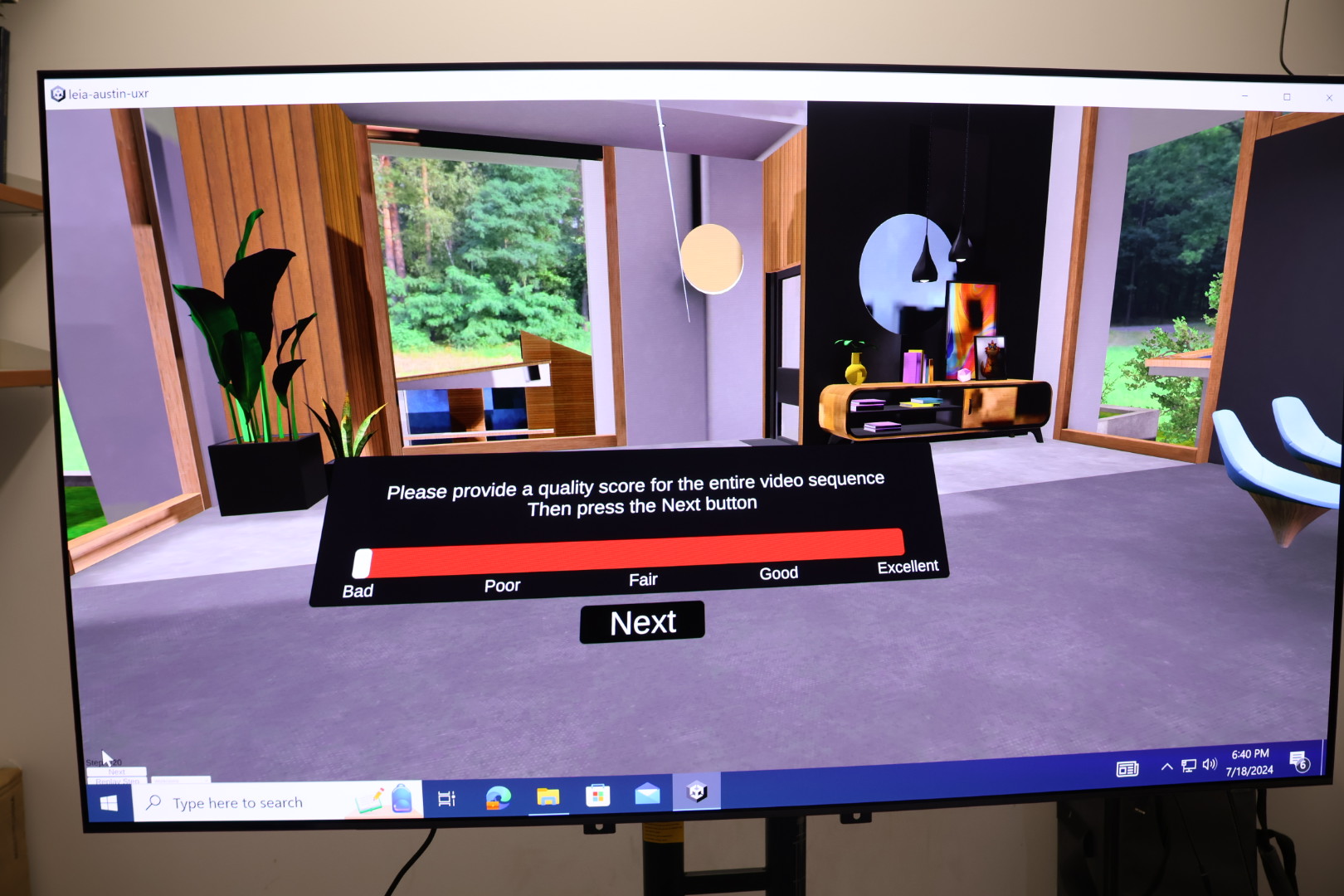}}
   \caption{(a) Initial state of the tool with a start button for users to begin the study. (b) Setup showing VR headsets, seating, and human avatar video playback on a monitor/TV. (c) Continuous rating bar displayed after each video for users to provide a quality score, with the cursor initially at the leftmost end.}
	\label{fig:environment}
\end{figure*}

\subsection{Subjective Study Environment}

The subjective quality assessment study was conducted in two separate rooms at the Laboratory of Image and Video Engineering at The University of Texas at Austin. The study utilized two Oculus Quest Pro headsets having resolutions of 1800$\times$1920 for each eye, using LCDs with a refresh rate of 72 Hz, and a field of view of 106 degrees horizontally and 96 degrees vertically. The participants used two controllers to interact with the human avatars during the study. The choice of the Oculus Quest Pro headset was based on its SOTA capabilities, relative affordability, and compatibility with many Metaverse applications, making it a suitable representative device presenting immersive experiences to human subjects.

The VR headsets were interfaced with two desktop computers equipped with an AMD Ryzen Threadripper PRO 5975WX 32-core CPU@3.60GHz, 256GB RAM, and two GeForce RTX 3090 Ti Graphics Cards. This setup allowed for simultaneous participation of two subjects. There was no distinguishable difference between the two desktop computers and the two Quest Pro headsets, ensuring a consistent experience between the two setups.

We have included photos of the testing environment to provide visual context for the subjective experiment setup. Fig. \ref{fig:initial-state} shows the initial state of the tool, where a start button allows users to begin the study. Fig. \ref{fig:video-playback} demonstrates the setup, including the arrangement of VR headsets, seating, and other equipment, with human avatar video playback on a monitor/TV. Fig. \ref{fig:rating-bar} illustrates the rating process, where a continuous rating bar is displayed after each video, allowing users to score the entire video sequence.

\subsection{Tool Design and Interface}\label{sec:tool}
Human avatar videos offer a suitable means to visualize objects and scenes within immersive applications that involve 6DoF. To enable subjects to observe human human avatar videos in a 6DoF environment, a privately released Unity tool provided by the Meta platform was utilized. The Unity tool, code will be made available, pending institutional approval, offered two modes of operation. The first mode allows live driving, enabling users to play human avatar videos while adjusting parameter configurations that control the appearance of artifacts. The second mode allows the playback of camera coordinates from a live HMD driving session, then dumps frames of ground truth and target. Users could monitor the tool using either a PC or a VR headset.

In the PC view mode, users had the freedom to explore a room and observe the human avatar from different viewing angles. The HMD provided observers with a synchronized display that accurately matched their body and head movements, creating a seamless and immersive perception of the virtual environment. In the VR view mode, a reminder window with a start button was provided to the users, allowing them to initiate the study using the controllers. The human avatar videos were then displayed, and after each video finished, a continuous rating bar was displayed, with a movable cursor initially positioned at the leftmost end. The quality bar was marked with five evenly-spaced Likert indicators, ranging from ``Bad" to ``Excellent." The ratings were then sampled as floating point numbers to one decimal place on [1, 5], with 1 indicating the lowest quality and 5 denoting the highest quality. Subjects adjusted the cursor position using the controller, then pressed the ``Next'' button to proceed to the next video sequence. The ratings were automatically recorded and saved in a CSV file. The application continued to play the subsequent videos in the playlist. Upon completion of all sequences, the subject was informed that the study was over.

To reduce any background distractions, a simple synthetic scene of a conference room was chosen and not varied across all the videos, with the exception of Video Group 1 (Table \ref{tab:group}), which had a different background as a control. This created a focused environment that allowed the participants to concentrate their efforts on providing accurate reports of the visual quality of the rendered avatar objects.

\begin{table}[!t]
\caption{Divisions of human subjects and videos into matched groups, ensuring balanced and unbiased viewing sessions.}\label{tab:group}
\centering
\normalsize
\resizebox{\columnwidth}{!}{%
\begin{tabular}{|c|c|c|c|}
\hline
Group & Participants & Session 1 & Session 2\\
\hline
I & 13 subjects & Subject Group A - Video Group 1 & Subject Group A - Video Group 2\\
\hline
II & 13 subjects & Subject Group A -  Video Group 2 & Subject Group A -  Video Group 1\\
\hline
III & 13 subjects & Subject Group B -  Video Group 3 &  Subject Group B -  Video Group 4\\
\hline
IV & 13 subjects & Subject Group B -  Video Group 4 & Subject Group B -   Video Group 3\\
\hline
V & 13 subjects & Subject Group C -  Video Group 5 & Subject Group C - Video Group 6\\
\hline
VI & 13 subjects & Subject Group C - Video Group 6 & Subject Group C - Video Group 5\\
\hline
\end{tabular}%
}
\end{table}

\subsection{Subjects and Training}\label{sec:subjects}

A total of 78 subjects (48 male and 30 female) from The University of Texas at Austin participated in the subjective human study. Their ages ranged from 18 to 33 years, with a majority falling between 20 and 25 years, as shown in Section \ref{sec:poststudy}. The participants had only limited or no familiarity with concepts of image and video processing. Some participants used only one of the two desktop computers, while others used both in two sessions. The computational power of the two PCs was very similar. The subjects were divided into six groups, and each participant completed two sessions as shown in Table \ref{tab:group}. As mentioned earlier, and explained in Table \ref{tab:content}, the videos were also divided into groups. Organizing the data in this way allowed every video to be viewed and rated by 26 subjects, but divided into two groups who viewed two video groups but in opposite order. Hence, Subject Groups I and II both viewed and rated Video Groups 1 and 2, but in reverse-ordered sessions. Similar protocol was applied for Subjects Group III and IV, and V and VI.

Prior to the first session, the subjects signed a consent form and received a general introduction to the study, informing them that they would be watching two different sets of videos in two separate sessions. The vision of each participant probed and recorded using the Snellen visual acuity test and the Ishihara color perception test, but no restrictions were placed on participation based on any visual deficiencies.

The subjects were asked to maintain a fixed sitting position at a distance of about 1.2 meters from the display. This distance aligns with the default separation between subjects and human avatars presented in HMDs. The prescribed distance aimed to afford participants adequate space to explore the human avatars from different angles, while also ensuring their comfort. The participants underwent a brief training session to acquaint them with the rating system and to provide instructions on how to assess and rate each video. The training session included six sequences generated from two contents, \textit{William Casual Talking} and \textit{Tony}, each presented at different qualities (pristine, slight distortion, severe distortion). These training sequences were not included in the test dataset.

During the testing sessions, a single-stimulus testing protocol was followed, as recommended by ITU-R BT 500.13 \cite{itutpic}. The ACR-HR methodology was employed to collect subjective scores, meaning that only one human avatar video was displayed during each rating. As mentioned in Section \ref {sec:adding_artifacts}, the reference videos used in this study were purchased from Metastage and reconstructed to closely resemble the original source human avatar videos. Although the capture device and reconstruction technology impose certain limitations, these reference videos are considered pristine for the purpose of our study. These were included in the human study, but the subjects were unaware of which videos were references. 

A randomized sequence of 120 human avatar videos, each with a duration of 14 or 15 seconds, was presented in each session. Consecutive videos from the same content were not played back-to-back to minimize visual memory effects. The participants were instructed to evaluate the perceived quality of the videos without considering aspects of the content, whether exciting, appealing, boring, or un/attractive. The randomized ordering aimed to minimize biases related to content preferences or relative ordering. Each subject spent approximately 5 to 10 seconds manipulating the cursor and rating each video, resulting in a total session time of 40 to 50 minutes. To avoid fatigue and biases, the subjects were instructed to participate in the second session more than 24 hours after the first.

\begin{table}[!t]
\caption{Participant feedback on the proportion of human avatar videos that elicited feelings of dizziness or discomfort. \label{tab:dizziness}}
\normalsize
\centering
\resizebox{\columnwidth}{!}{
\begin{tabular}{|c|c|c|c|c|c|c|c|}
\hline
\makecell[c]{\# of human avatar\\ videos inducing\\ uneasiness/dizziness} & None & $<$10 & 10-19 & 20-39 & 40-100 & $>$100 & \makecell[c]{Didn't\\ remember\\ when}\\
\hline
\% of sessions & 56\% & 6.7\% & 4\% & 8.1\% & 2.7\% & 4.7\% & 17.5\% \\
\hline
\end{tabular}%
}
\end{table}

\begin{figure}[!t]
	\centering
	\normalsize
	\renewcommand{\tabcolsep}{1pt} 
	\renewcommand{\arraystretch}{1} 
	\captionsetup[subfloat]{labelfont=footnotesize,textfont=footnotesize}
    \subfloat[]{\includegraphics[width=\columnwidth]{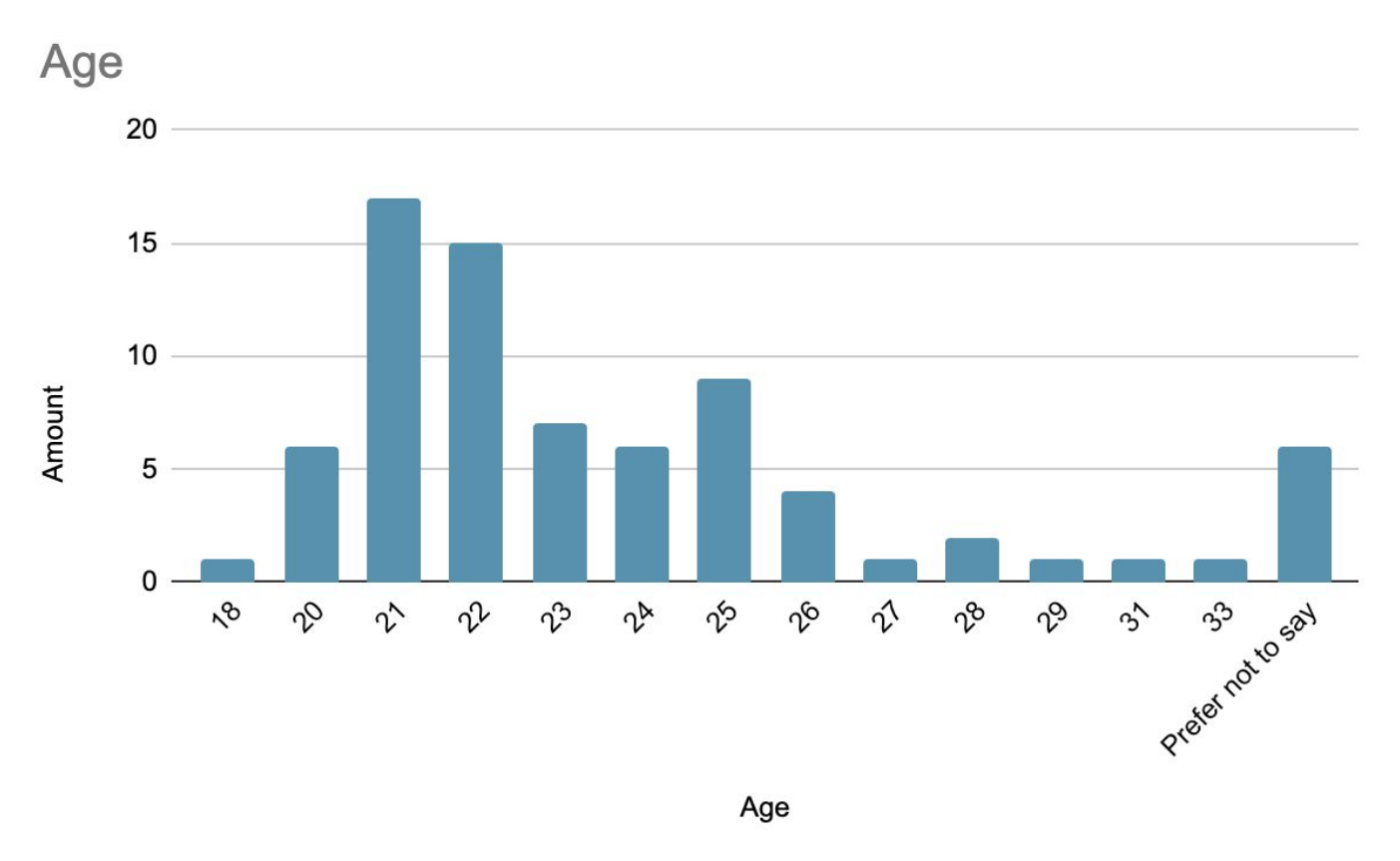}}\\
     \subfloat[]{\includegraphics[width=\columnwidth]{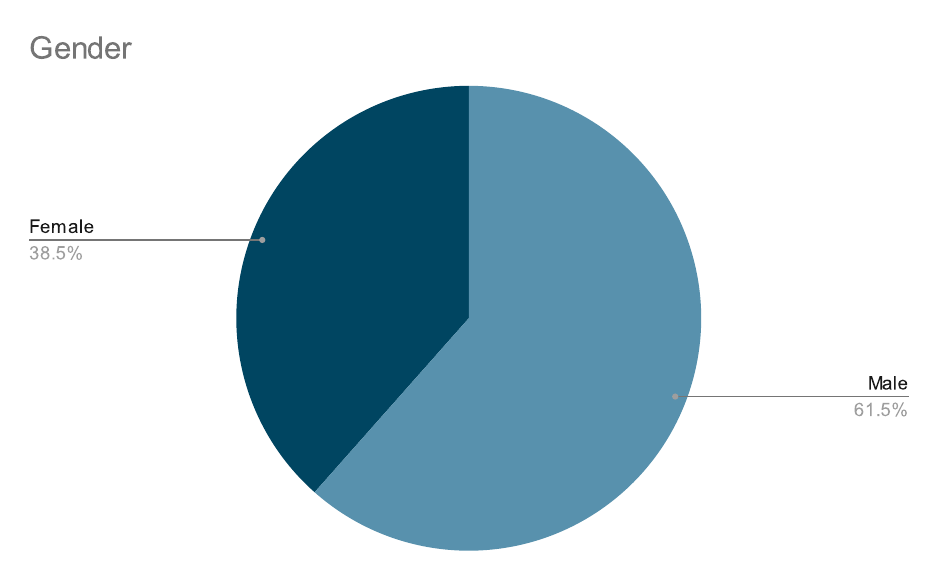}}
   \caption{Demographics of human study participants based on age (a) and gender (b).}
	\label{fig:demograph}
\end{figure}

\subsection{Post Study Questionnaire} \label{sec:poststudy}

After completing each session, each participant was requested to provide feedback on the study via a post-survey questionnaire. This subsection offers an outline of the post-survey questionnaire and presents relevant demographic information pertaining to the participants.

To assess the adequacy of the durations of the displayed videos that the subjects rated the quality of, a specific question was included. Among the 156 sessions conducted (comprising 78 subjects, two sessions per subject), the subjects reported that in 155 sessions (99.3\%) that the 15-second duration was sufficient to rate overall video quality. Another question examined the perception of the overall distribution of the video quality of the videos. In 128 sessions (81.8\%), the participants reported that the distribution was uniform, indicating an equal representation of videos across quality levels. However, in some sessions the participants generally perceived the majority of the videos to be either of above average quality or of below average quality.

To evaluate the difficulty experienced by the subjects when rating the perceptual quality of the videos, another question requested them to rate the difficulty (after each session) on a scale of 0 to 5, with 0 indicating very difficult and 5 indicating reasonably easy to make quality judgments. Among the 156 sessions, only one session was reported as generally difficult to make subjective quality ratings, indicating that the majority of participants were able to rate subjective quality without encountering significant difficulty. This observation is supported by the mean difficulty score of 3.48 and the median difficulty score of 4, indicating only a moderate level of difficulty.

Participants were also queried about any experiences of dizziness or uneasiness during the viewing and rating of the videos. Approximately 44\% of sessions reported some degree of dizziness or uneasiness, but this generally occurred only on a small percentage of the videos, as shown in Table \ref{tab:dizziness}. Generally, only a very small number of videos elicited these feelings, and none caused any of the subjects to stop their participation.

At the conclusion of each session, demographic information, including age and gender, was collected. The mean age of the participants was 23.03, with a median age of 22 and a standard deviation of 2.69.  Visualizations depicting the age and gender distributions are provided in Fig. \ref{fig:demograph}.

\begin{table}[!t]
\caption{Subject consistency scores (SRCC and PLCC) for six video groups, indicating high levels of inter-subject and intra-subject reliability.}\label{tab:consistency}
\small
\centering
\resizebox{0.8\columnwidth}{!}{%
\begin{tabular}{|c|p{1.6cm}<{\centering}|p{1.6cm}<{\centering}|p{1.6cm}<{\centering}|p{1.6cm}<{\centering}|}
\hline
 & \multicolumn{2}{c|}{Inter-Subject Consistency} & \multicolumn{2}{c|}{Intra-Subject Consistency}\\
\hline
Video Group & SRCC & PLCC & SRCC & PLCC\\
\hline
1 & 0.9420 & 0.9748 & 0.8444 & 0.8833\\
\hline
2 & 0.9524 & 0.9798 & 0.8729 & 0.9084\\
\hline
3 & 0.9611 & 0.9763 & 0.8759 & 0.8809 \\
\hline
4 & 0.9591 & 0.9780 & 0.8706 & 0.8855 \\
\hline
5 & 0.9382 & 0.9711 & 0.8353 & 0.8774 \\
\hline
6 & 0.9436 & 0.9736 & 0.8363 & 0.8726 \\
\hline
\end{tabular}%
}
\end{table}

\begin{table}[!t]
\caption{Subject consistency (SRCC and PLCC) before and after changing the background, showing minimal impact on video quality perception.} \label{tab:background}
\centering
\footnotesize
\resizebox{0.8\columnwidth}{!}{
\begin{tabular}{|c|p{1.6cm}<{\centering}|p{1.6cm}<{\centering}|}
\hline
 & \multicolumn{2}{c|}{Inter-Subject Consistency}\\
\hline
Video Group & SRCC & PLCC\\
\hline
1 (before: 12 ratings / after: 15 ratings) & 0.9293 & 0.9746\\
\hline
2 (before: 13 ratings / after: 14 ratings)  & 0.9590 & 0.9804\\
\hline
\end{tabular}%
}
\end{table}
\subsection{Subject-Consistency Analysis} \label{sec:subject-consistency-analysis}

To study the internal consistency of the collected data, we analyzed inter-subject and intra-subject correlations of the raw data collected from the participants. As previously mentioned, the 78 participants were evenly divided into six groups, as outlined in Table \ref{tab:content}. To compute the inter-subject consistency scores, each video group's subject ratings of every video were divided into two separate and non-overlapping subsets of equal size. This procedure was repeated 100 times with random splits. The median values of the Spearman's rank-ordered correlation coefficient (SRCC) and the Pearson linear correlation coefficient (PLCC) between the Mean Opinion Scores (MOS) of the two subsets were computed and are presented in Table \ref{tab:consistency}. Across all subject groups, the average SRCC and PLCC values representing inter-subject consistency were determined to be 0.9494 and 0.9756, respectively.

Intra-subject consistency measurements were also calculated to assess the level of consistency exhibited by the individual subjects when rating the videos. For each subject group, the SRCC and PLCC were calculated between the individual opinion scores and MOS. This procedure was repeated for all 78 subjects across all subject groups. Table \ref{tab:consistency} presents the median SRCC and PLCC values for each subject group. The overall average SRCC and PLCC across all subject groups were determined to be 0.8559 and 0.8847, respectively. These scores encourage a high level of confidence in the acquired opinion scores.

To further investigate the reliability of the data, an inter-subject consistency check was conducted on Video Groups 1 and 2 to determine whether changing the background influenced the subjects' ratings and to assess the consistency of their responses. The SRCC and PLCC between the two sets of ratings were computed between the two sets of ratings for each video group. The results of this analysis are presented in Table \ref{tab:background}. For Video Group 1, which had the different background, the SRCC and PLCC values between the two sets of ratings were found to be 0.9293 and 0.9746, respectively. In the case of Video Group 2, characterized by a distinct background, the ratings exhibited a strong positive correlation before and after the background change, as indicated by the high SRCC (0.9590) and PLCC (0.9804) values. The high correlation scores between the two sets of ratings for both video groups suggest that changing the background did not significantly impact the subjects' perception of video quality.

The analysis of subject consistency also revealed strong agreement among the subjects, both within subjects and across different backgrounds, lending credibility to the subjective ratings collected in this study.

\begin{figure*}[htbp]
    \centering
    \captionsetup[subfloat]{labelfont=footnotesize,textfont=footnotesize}
    \subfloat[Recovered quality scores using SUREAL.\label{fig:recovered_score}]{
    \includegraphics[width=\textwidth]{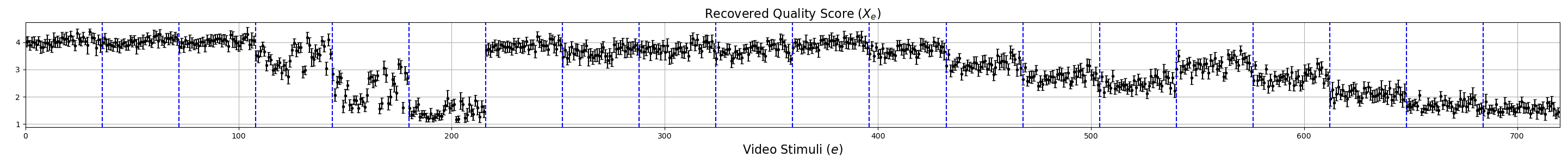}}\\
    \subfloat[\label{fig:recoverd_subject}]{
    \includegraphics[width=.485\textwidth]{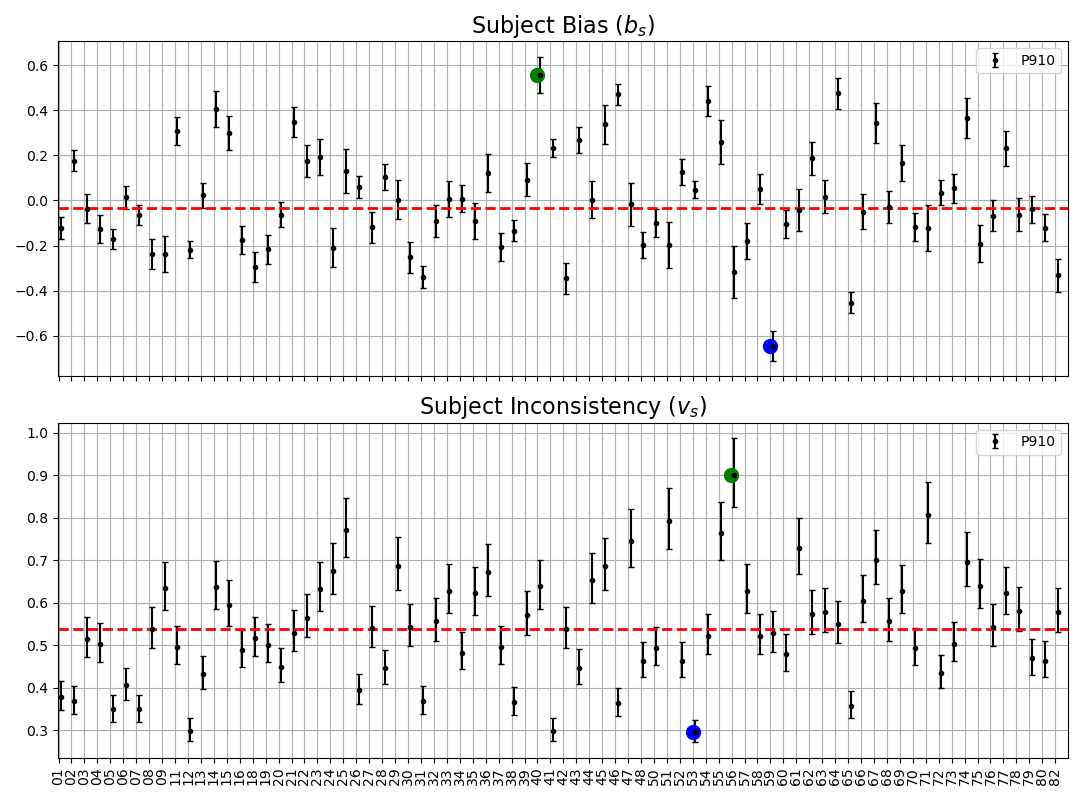}}\hspace{5pt}
    \subfloat[\label{fig:recovered_content}]{\includegraphics[width=.485\textwidth]{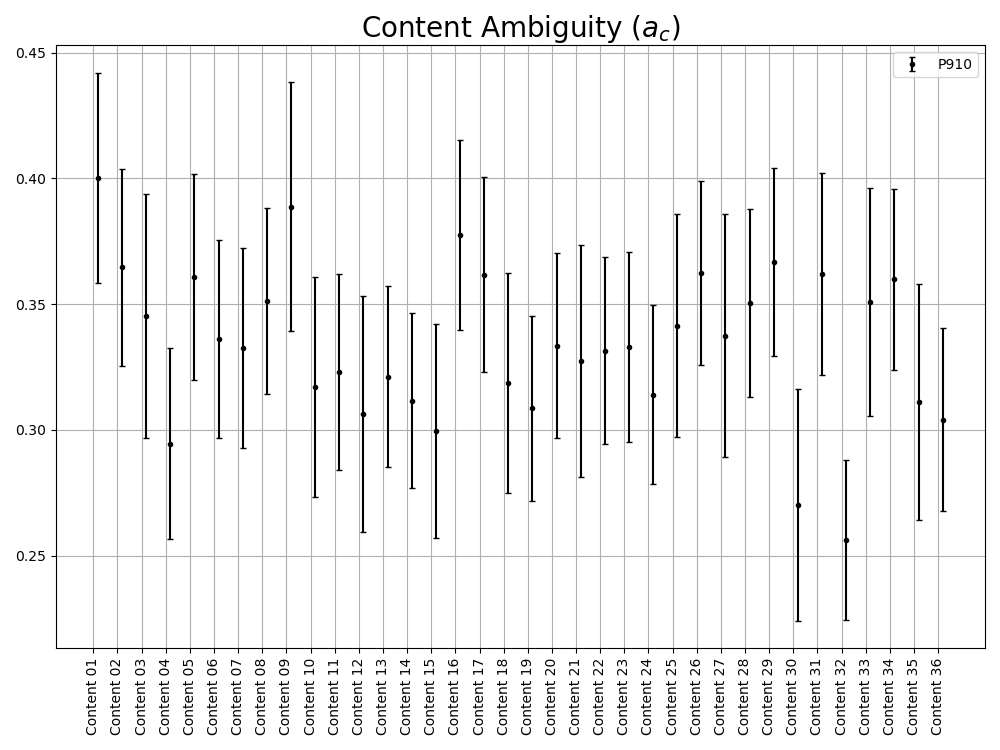}}\\
    \caption{The MLE formulation provides (a) estimated final opinion scores, along with (b) subject bias and inconsistency figures and (c) content ambiguity. Each of these results includes the estimated parameters and their 95\% confidence intervals. The content index shown in (c) corresponds to the descriptions listed in Table \ref{tab:content}}%
    \label{fig:recovered}
\end{figure*}

\begin{figure}[!t]
  \centering
  \footnotesize
  \renewcommand{\tabcolsep}{1.3pt} 
  \renewcommand{\arraystretch}{1.3} 
  \captionsetup[subfloat]{labelfont=footnotesize,textfont=footnotesize}
  \subfloat[\label{fig:MLE-MOS}]{\includegraphics[height=120pt]{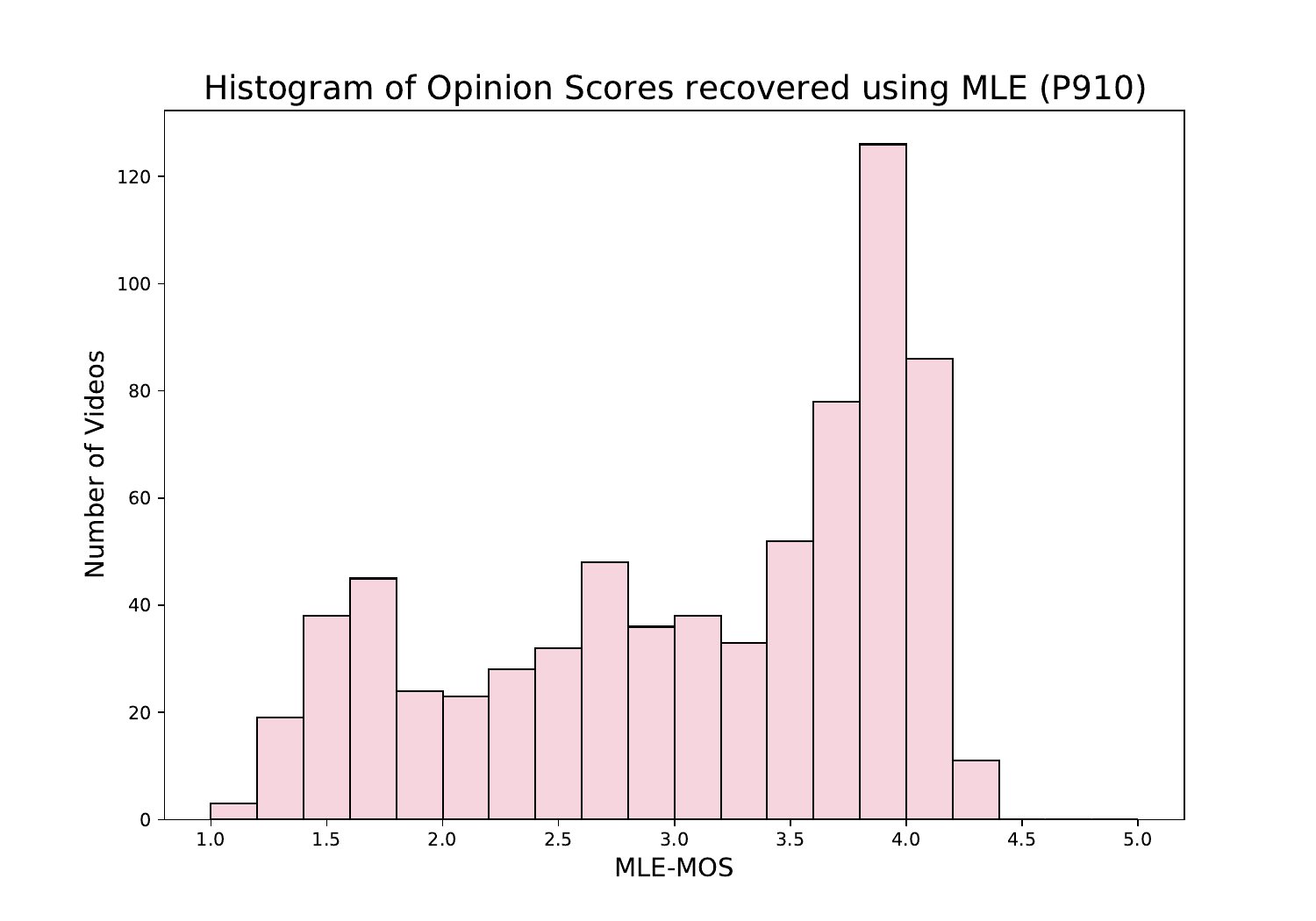}}\\
  \subfloat[\label{fig:DMOS}]{\includegraphics[height=120pt]{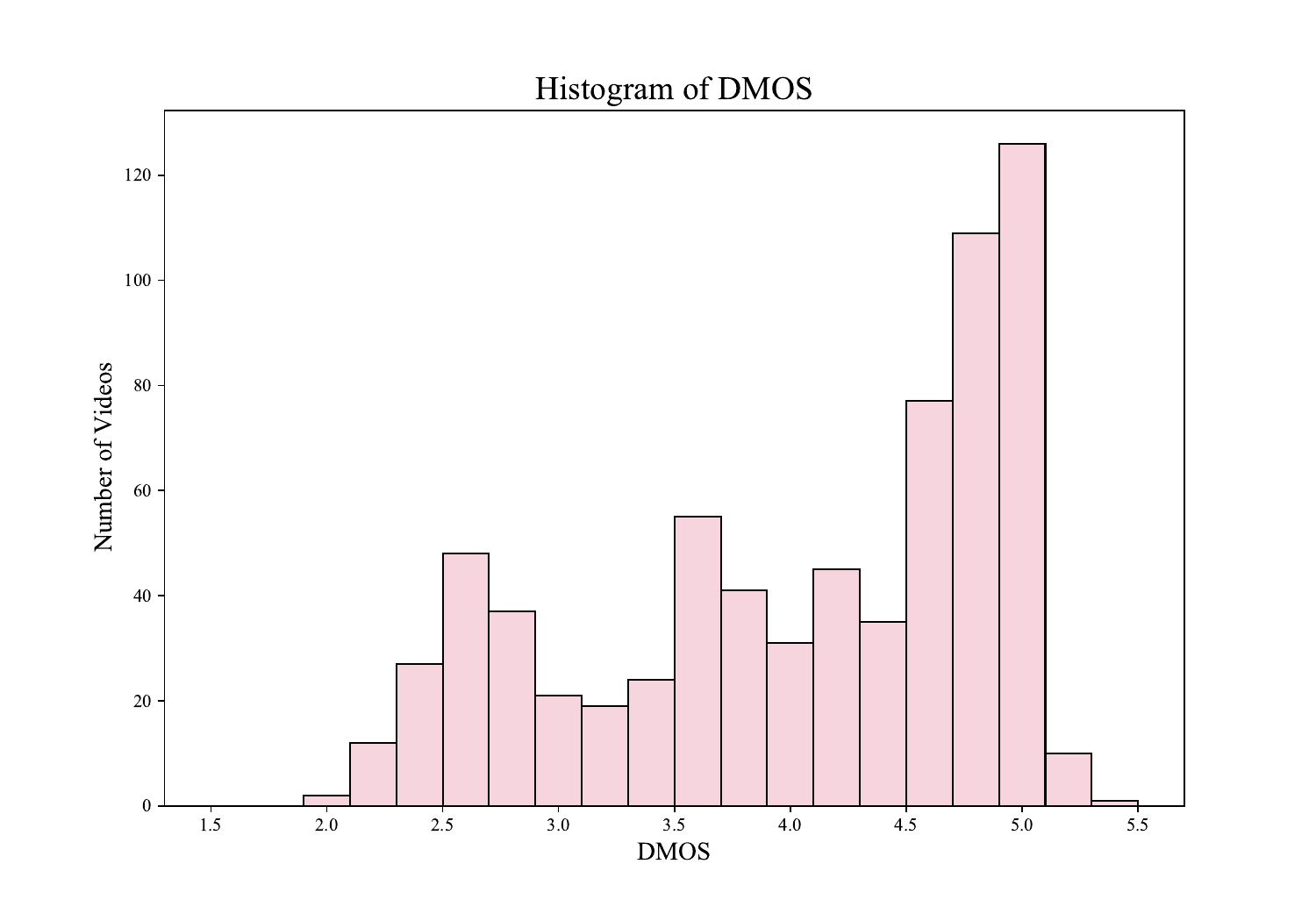}}\\
  \caption{Histograms of (a) MLE-MOS and (b) DMOS from the LIVE-Meta Rendered Human Avatar VQA Database. The histogram is partly but incompletely right-skewed, indicating that while most videos are of good quality, very few are of excellent quality.} 
\end{figure}

\subsection{Processing of the Subjective Scores}

We applied the SUREAL MLE-MOS method \cite{7921900} to recover reliable subjective quality scores on the human avatar videos using the P.910 model \cite{li2020simple}. This method utilizes a maximum likelihood estimate (MLE) approach to compute MOS, offering advantages over prior subject rejection protocols \cite{itutpic, 5404314}. MLE-MOS stands for Maximum Likelihood Estimate - Mean Opinion Score, a statistical method used to derive the most likely subjective quality scores from the data collected in the study.

Subjective experiments are known to have various issues around the reliability and accuracy of individual scores, necessitating some level of statistical analysis to provide better mean opinion scores. For example, in earlier versions of the P.910 standard \cite{itutpic}, certain opinion scores are excluded as outliers when they deviate significantly from the average. MLE-MOS offers a better alternative, where each subject is characterized by a bias and inconsistency term, roughly reflecting how much this rater's scores are lower/higher than the average, and how much they spread around the global average. By correcting each rater's scores for the estimated bias and weighting them inversely proportional to its variance, one obtains a maximum-likelihood estimate of the mean opinion score that is both more accurate and offers tighter confidence intervals. Additionally, this method allows the estimation of bias and inconsistency per rater, providing valuable insights into video contents and test subjects.

The MLE-MOS iterative algorithm that estimates all these parameters has been implemented in the open-source SUREAL software package \cite{7921900} and has been used in our experiments.Our choice of the SUREAL model is motivated by several factors: its decreased susceptibility to subject bias, the provision of narrower confidence intervals, robust handling of missing data, and the ability to offer detailed insights into video contents and test subjects. These aspects are crucial for maintaining the integrity and precision of subjective quality assessments, ensuring that the analysis remains reliable even with incomplete datasets.

We will refer to MOS processed in this way as MLE-MOS. These include decreased susceptibility to subject bias, narrower confidence intervals, robust handling of missing data, and the ability to provide insights into video contents and test subjects.

In SUREAL, the raw video ratings are represented as random variables ${{X_{e,s}}}$:
\begin{gather}
    \label{eq:sureal}
    X_{e,s} = x_e + B_{e,s} + A_{e,s},  \\   
    B_{e,s} \sim \mathcal{N}(b_s,v_s^2), \notag\\
    A_{e,s} \sim \mathcal{N}(0,a_{c:c(e)=c}^2). \notag
\end{gather}
for $e=1,2,3,...,720$ and $s=1,2,3,...,78$. Let $x_e$ denote the perceived quality of video $e$ as assessed by an impartial and consistent hypothetical viewer. The bias ($b_s$) and inconsistency ($v_s^2$) associated with human subject $s$ are represented by i.i.d. Gaussian variables $B_{e,s}$. Assume that the bias and inconsistency of human subjects are consistent across all rated videos. The ambiguity ($a^2_c$) associated with a specific video content $c$ is modeled by i.i.d. Gaussian variables $A_{e,s}$. In this database, the unique source sequences are denoted as $c=1,2,...,36$. Also assume that all distorted videos from same video source exhibit a uniform level of ambiguity, and that this ambiguity remains consistent across all users and video content. To estimate the parameters $\theta=({x_e},{b_s},{v_s},{a_c})$ which represent the variables of the model, MLE is utilized, and the log-likelihood function $L$ is formulated as follows:
\begin{equation}
    L = \log P(\{x_{e,s}\}|\theta). 
\end{equation} 
Estimation of the optimal solution $\hat\theta = \arg \max_{\theta} L$ is performed using data collected from the psychometric study, utilizing the Belief Propagation algorithm described in \cite{7921900}.

Figure \ref{fig:recovered} visually illustrates the estimated parameters. Figure \ref{fig:recovered_score} plots the recovered quality scores of the 720 videos in the database created with the 20 different distortion parameter settings listed in Table \ref{tab:distortion}. Since every set of 36 videos corresponds to the same distortion type, we have added vertical lines to distinguish the distortion types. As expected, the mean predicted video quality scores noticeably decline as the color resolution is decreased. The results highlight that color resolution and frame rate had a more significant impact on the perceived quality of human avatar videos than did delay variations. This suggests that optimizing delay settings could help achieve data efficiency in VR human avatar video streaming without compromising perceptual quality.

As depicted in Fig. \ref{fig:recoverd_subject}, our analysis of subject bias and inconsistency on human avatar videos reveals significant individual differences in perception and evaluation standards among the subjects. From the parameter estimates, subject \#58 had the bias value $b_s = -0.64$, the lowest among all the subjects, indicating that their quality scores tended to be lower than those of the other subjects. Conversely, subject \#39 had the highest bias value ($b_s = 0.55$), suggesting that their quality scores tended to be higher. The median bias value obtained was -0.03, reflecting a moderate overall bias. In terms of inconsistency, subject \#55 exhibited the highest value ($v_s = 0.89$), indicating greater variability of their quality scores, while subject \#52 exhibited the lowest inconsistency ($v_s = 0.29$). The median inconsistency across subjects was 0.53. These insights underscore the importance of accounting for individual differences amongst subjects. In Fig. \ref{fig:recoverd_subject}, the subjects having the highest and lowest biases, and the subjects having the highest and lowest inconsistencies are highlighted by superimposed green and blue dots. Additionally, the red dotted horizontal lines indicate the median values. By identifying and adjusting for these biases and inconsistencies, it is possible to enhance the robustness and accuracy of the quality ratings. This refined understanding can inform the weighting of subject ratings and/or the exclusion of outliers.

The level of ambiguity of the 36 source human avatar videos is depicted in Fig. \ref{fig:recovered_content}. Among these videos, the human avatar video from the \textit{01-Natasha Serious Talking} content exhibited the highest ambiguity with a value of $a_c = 0.40$, whereas the human avatar video from the \textit{32-Jenny Sitting Casual} content displayed the lowest ambiguity with a value of $a_c = 0.25$. This indicates that some contents were more challenging for viewers to evaluate consistently, likely due to variations in movement, expression, or other factors that affect perceptual quality.

MLE-MOS has established itself as a dependable subjective data processing protocol, especially in scenarios where reference pristine videos are absent, making it particularly valuable in the advancement and assessment of NR VQA algorithms. Conversely, Differential Mean Opinion Scores (DMOS) are commonly employed in the development of FR VQA algorithms, reducing the dependence of quality labels on the content. Here, the original human avatar videos provided by Metastage serve as proxy reference videos for the calculation of DMOS. The DMOS of the $i^{th}$ video sequence is determined as:
\begin{equation} \label{eq:dmos}
DMOS(i) = 5 - (MOS(ref(i)) - MOS(i)).
\end{equation}

In this context, MOS($i$) denotes the $i^{th}$ video that has undergone distortion, determined via the MLE formulation. The proxy reference video is denoted as ref($i$).

\begin{figure*}[!htbp]
    \centering
    \captionsetup[subfloat]{labelfont=footnotesize,textfont=footnotesize}
    \renewcommand{\tabcolsep}{1.3pt} 
    \renewcommand{\arraystretch}{1.3} 
    \subfloat[Videos with odd IDs were assigned distorted videos with delays of 100 ms and 300 ms\label{fig:delay_odd}]{\includegraphics[width=.45\textwidth]{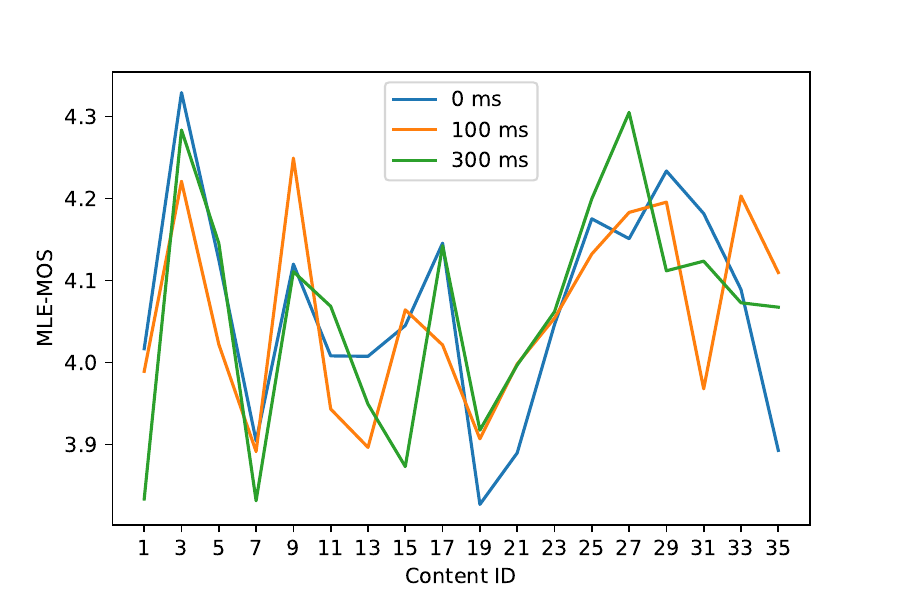}}\hspace{5pt}
    \subfloat[Videos with even IDs were assigned distorted videos with delays of 200 ms and 400 ms\label{fig:delay_even}]{\includegraphics[width=.45\textwidth]{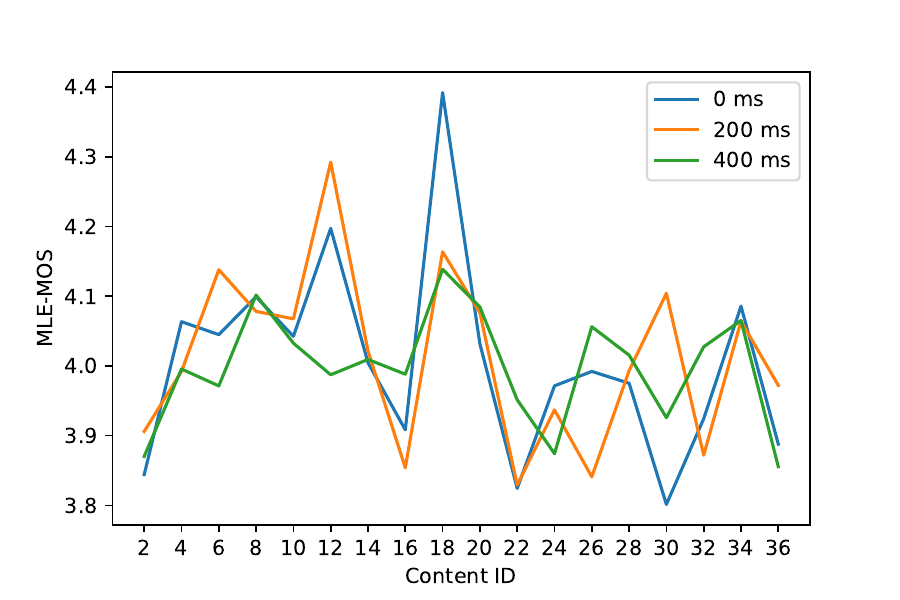}}\\
    \subfloat[Videos with odd IDs were assigned distorted videos with color resolutions of 1600p, 1080p, and 640p\label{fig:color_resolution_odd}]{\includegraphics[width=.45\textwidth]{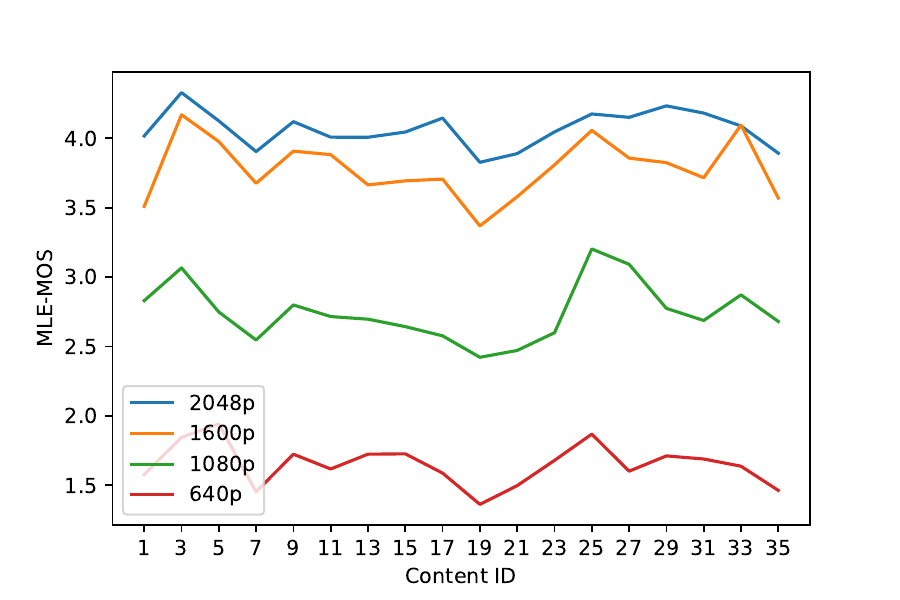}}\hspace{5pt}
    \subfloat[Videos with even IDs were assigned distorted videos with color resolutions of 1280p, 720p, and 480p\label{fig:color_resolution_even}]{\includegraphics[width=.45\textwidth]{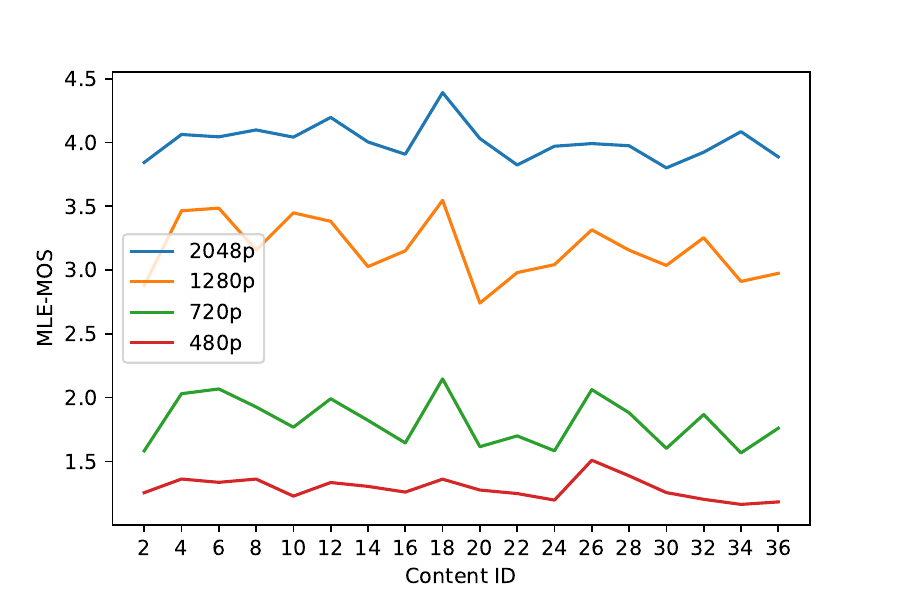}}\\
    \subfloat[Rate distortion curves for video with different frame rates = 30fps, 20fps, 10fps, and 15$\pm$10 fps\label{fig:fps}]{\includegraphics[width=\textwidth]{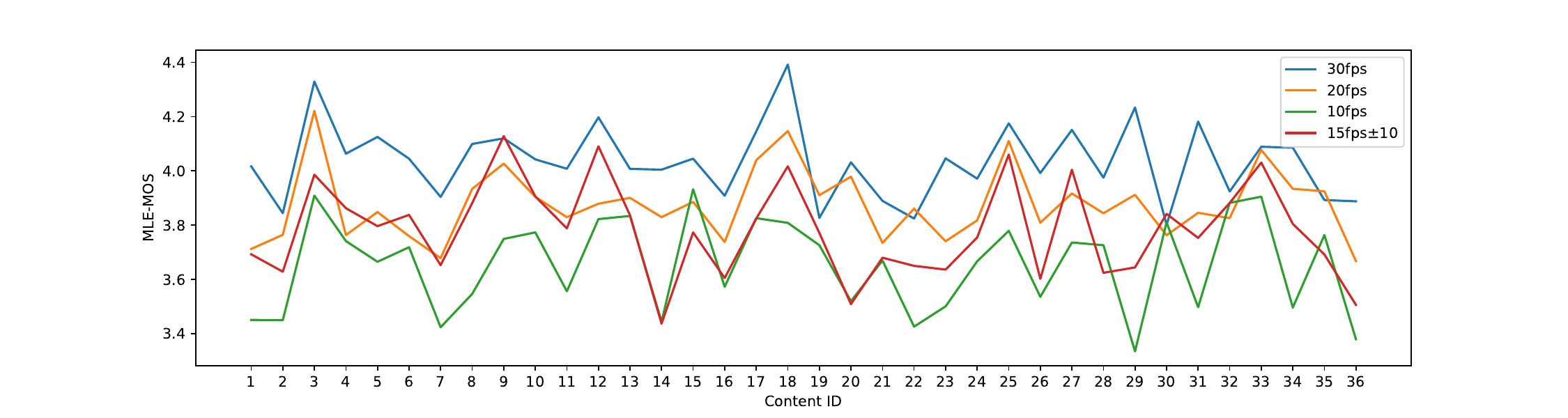}}\\
    \subfloat[Variation of MLE-MOS with distortion index across contents with odd IDs and even IDs\label{fig:dis_type}]{\includegraphics[width=\textwidth]{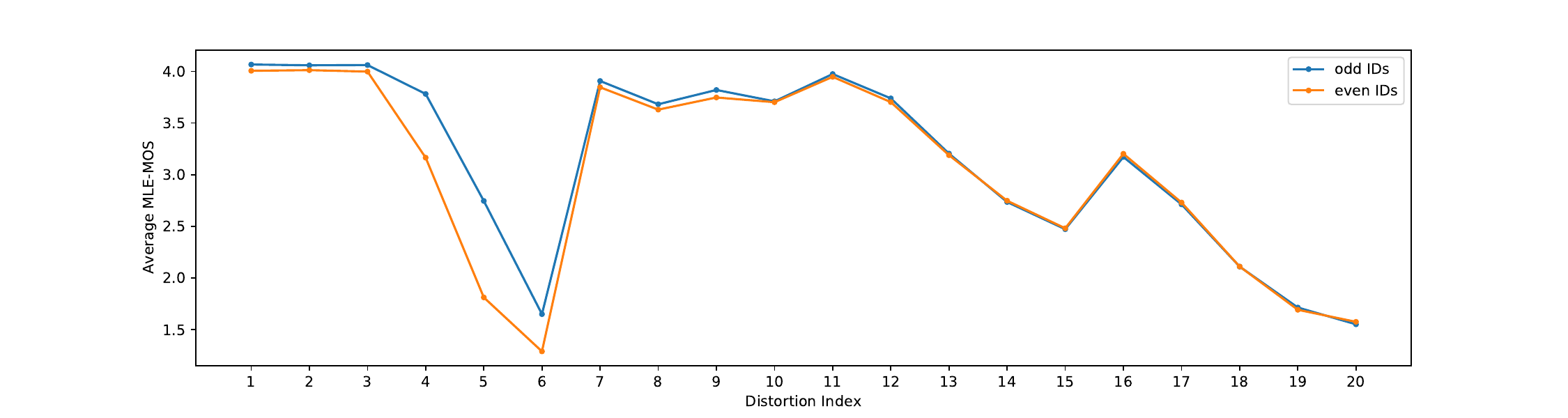}}\\
   \caption{(a)-(e) Variation of MLE-MOS against content for varying delays, color resolutions, and frame rates, (f) Variation of MLE-MOS with distortion index across contents with odd IDs and even IDs \label{fig:against_content}}
\end{figure*}

\subsection{Data Analysis}

Figure \ref{fig:MLE-MOS} displays the MLE-MOS histogram obtained using SUREAL. The MLE-MOS values spanned the range [1.162, 4.391]. The distribution exhibits a slight right-skew, consistent with patterns observed in other VQA databases. The majority of the videos are of good quality, but few videos fall into the category of excellent quality. The histogram in Figure \ref{fig:DMOS} plots the distribution of DMOS calculated using Equation \ref{eq:dmos}. The DMOS values spanned the range [1.968, 5.302]. The distribution of the DMOS closely resembles that of the MLE-MOS, albeit with a slight rightward shift.

\subsubsection{Impact of Delays on MLE-MOS}
To investigate the impact of delays on MLE-MOS, we analyzed the MLE-MOS of videos with different content IDs for various delay values. Fig. \ref{fig:delay_odd} presents the MLE-MOS curves for videos with odd content IDs, which were assigned distorted videos with delays of 100 ms and 300 ms. Fig. \ref{fig:delay_even} displays the MLE-MOS curves for videos with even content IDs, which were assigned distorted videos with delays of 200 ms and 400 ms.

The grouping of content IDs into odd and even categories was done to systematically assess the impact of different delay values on perceived video quality. By alternating the delay values between odd and even content IDs, we aimed to ensure a balanced and comprehensive evaluation across the dataset.

It can be observed that there are no significant separation between curves for MLE-MOS across different delay values. This suggests that the delay parameter has minimal influence on the perceived quality of the videos.

\subsubsection{MLE-MOS Content Dependence}
In order to investigate the relationship between the source video contents, single distortions of color resolution, and their combined impact on MLE-MOS, we analyzed the results presented in Fig. \ref{fig:color_resolution_odd} and Fig. \ref{fig:color_resolution_even}. Fig. \ref{fig:color_resolution_odd} illustrates the MLE-MOS curves corresponding to videos with odd content IDs, which were assigned distorted videos with color resolutions of 1600p, 1080p, and 640p. Fig. \ref{fig:color_resolution_even} shows those with even content IDs, which were assigned distorted videos with color resolutions of 1280p, 720p, and 480p. These figures depict distinct demarcations between the MLE-MOS curves associated with different color resolutions, highlighting the influence of color resolution on perceptual quality assessment.

\subsubsection{Rate Distortion Curves}
The impact of varying frame rates on MLE-MOS was examined by analyzing the MLE-MOS curves shown in Fig. \ref{fig:fps}. Notably, distinct separations can be observed between curves of MLE-MOS of videos with frame rates of 30 fps, 20 fps, and 10 fps. However, the curves for videos with a frame rate of 15$\pm$10 fps exhibit inconsistent placement. At times, these curves fell between the curves of 20 fps and 10fps, while in other instances, they surpassed the 20 fps curve or align closely with the 10fps curve. This finding indicates that the introduction of variance in frame rate across different content may hinder individuals' ability to perceive changes in perceptual quality, as the content may consist of varying levels of static or dynamic movements. The observed decrease in MLE-MOS values at a lower frame rate (10 fps) highlights the significant influence of temporal distortions on the perceived quality of human avatar videos.

\subsubsection{Distortion Type Analysis}

Figure \ref{fig:dis_type} provides an analysis of the influence of distortion type on the MLE-MOS across different content IDs. It may be observed that the curves for contents with odd and even IDs exhibit consistency, except for a distinct difference observed in the curves corresponding to distortion indices 4, 5, and 6, which represent color resolution distortions. As discussed in Section III-B of the Main paper, video content with odd IDs was assigned color resolution distortions of 1600p, 1080p, and 640p, while video content with even IDs was assigned resolutions of 1280p, 720p, and 480p, respectively. Thus, the curves depicted in Figure \ref{fig:dis_type} effectively reflect this assignment strategy and its impact on the MLE-MOS.

These analyses highlight several key findings regarding the impact of different factors on the visual perception of video quality, particularly in the context of human avatar videos. Among these factors, color resolution and varying frame rate had a stronger impact on video quality perception of human avatar videos compared to different delay values. These findings suggest that adjusting delay can lead to data efficiencies in VR human avatar video streaming without significant perceptual loss. For VR content creators, understanding which distortions have the least impact on perceived quality can help optimize encoding and streaming workflows. For instance, reducing frame rates moderately may save bandwidth without greatly affecting user experience, while maintaining high color resolution could avoid noticeable degradation in quality. Monitoring these kinds of balances can lead to more efficient use of resources while maintaining an acceptable level of quality.

\section{Evaluation framework} \label{evaluation}

To show the usefulness of the new subjective database, we conducted a comparative study of existing VQA models on it. Next we detail the processes of data set preparation, the model evaluation protocol, and the performance outcomes of the compared FR and NR IQA/VQA models.

\begin{table*}[!t]
   \renewcommand{\arraystretch}{1.3}
   \caption{Median SRCC, KRCC, PLCC and RMSE of the compared FR video quality models against human judgments of quality of the videos in the LIVE-Meta Rendered Human Avatar VQA database over 1000-train-test splits. The \underline{underlined} and \textbf{boldfaced} items represent the best and top three performers, respectively. The contrique (fr), re-iqa (fr), conviqt (ft), and holoqa models achieved standout performance.}
   \label{tab:FRcomparison}
   \setlength{\tabcolsep}{3pt}
   \centering
   \renewcommand{\tabcolsep}{3.3pt} 
   \resizebox{\textwidth}{!}{
   \begin{tabular}{|c|c|cccc|cccc|}
   \hline
   \multirow{2}{*}{Metrics} & \multirow{2}{*}{Pre-Training / Fine-tuning Dataset} & \multicolumn{4}{c|}{Body-only} & \multicolumn{4}{c|}{Face-only}\\
   \cline{3-6}\cline{7-10}
   & & SRCC($\uparrow$) & KRCC($\uparrow$) & PLCC($\uparrow$) & RMSE($\downarrow$)& SRCC($\uparrow$) & KRCC($\uparrow$) & PLCC($\uparrow$) & RMSE($\downarrow$) \\
   \hline
   PSNR-RGB~  & N/A (\textit{handcrafted}) & ~0.7219~ & ~0.5223~ & ~0.7562~ & 0.5750~ & 0.7129 & 0.5111 & 0.7405 & 0.5924\\
   PSNR-Y~ & N/A (\textit{handcrafted}) & 0.7167 & 0.5192 & 0.7505 & 0.5798 & 0.6983 & 0.4987 & 0.7285 & 0.6026\\
   SSIM, 2004 \cite{1284395} & N/A (\textit{handcrafted}) &  0.7694 & 0.5618 & 0.7755& 0.5537 & 0.8001 & 0.5929 & 0.8264 & 0.4917\\
   DLM, 2011 \cite{5765502} & N/A (\textit{handcrafted}) & 0.8599 & 0.6601 & 0.8919 & 0.3995 & 0.8716 & 0.6762 & 0.9106 & 0.3624\\
   VIF, 2005 \cite{1532311, 1576816} & N/A (\textit{handcrafted}) & 0.7184 & 0.5200 & 0.7499 & 0.5819 & 0.8154 & 0.6085 & 0.8513 & 0.4621\\
   LPIPS (AlexNet), 2018 \cite{Zhang_2018_CVPR}~  & ImageNet pretrained AlexNet  & -0.8616 & -0.6573 & 0.8815 & 0.4131 & -0.8790 & -0.6835 & 0.9069 & 0.3703\\
   LPIPS (VGG), 2018 \cite{Zhang_2018_CVPR}~  & ImageNet pretrained VGG-16   & -0.8441 & -0.6356 & 0.8535 & 0.4558 & -0.8764 & -0.6803 & 0.9047 & 0.3734\\
   CONTRIQUE (FR), 2022 \cite{9796010} & ImageNet pretrained ResNet-50 / KADIS-700k + AVA + COCO + CERTH-Blur + VOC & 0.9024 & 0.7205 & 0.9470 & 0.2816 & \textbf{0.8991} & 0.7138 & 0.9403 & \textbf{0.2976}\\
   Re-IQA (FR), 2023 \cite{Saha_2023_CVPR} & ImageNet pretrained ResNet-50 \& ImageNet pretrained MoCo-v2 / KADIS-140K + AVA + COCO + CERTH-Blur + VOC & \textbf{0.9068} &\textbf{0.7298} & 0.9401 &0.2919 & 0.8988 & \textbf{0.7154} & 0.9308 & 0.3218\\
   \hline
   VMAF (v0.6.1), 2016 \cite{li2016toward}~  & N/A (\textit{handcrafted}) & ~0.8179~ & ~0.6085~ & ~0.8367~  & ~0.4789~ & 0.8695 & 0.6716 & 0.8949 & 0.3905 \\
   FovVideoVDP (v1.2.0), 2021 \cite{mantiuk2021fovvideovdp}~  & N/A (\textit{handcrafted}) &  0.7763 & 0.5750 & 0.8103 & 0.5168 & 0.8810 & 0.6888 & 0.9155 & 0.3527  \\
   ST-GREED, 2021 \cite{Madhusudana_2021} & N/A (\textit{handcrafted}) & 0.8543 & 0.6619 & 0.8800 & 0.4162 & 0.8946 & \textbf{0.7154} & 0.9148 & 0.3552  \\
   FUNQUE, 2022 \cite{9897312} & N/A (\textit{handcrafted}) & 0.8655 & 0.6684 & 0.9049 & 0.3740 & 0.8980 & 0.7140 & \textbf{0.9404} & 0.2982  \\
   Y-FUNQUE+, 2023 \cite{10375336} & N/A (\textit{handcrafted}) & 0.8663 & 0.6641 & 0.8914 & 0.3971 & 0.8956 &0.7120 & 0.9355 & 0.3094  \\
   3C-FUNQUE+, 2023 \cite{10375336} & N/A (\textit{handcrafted}) & 0.8599 & 0.6547 & 0.8832 & 0.4104 & 0.8981 & 0.7129 & 0.9357 & 0.3101  \\
   CONVIQT (FR), 2023 \cite{10243566} & ImageNet pretrained ResNet-50 + Kinetics-400 + Waterloo1k + dareful + REDS + MCML + UVG & 0.9065 & 0.7289 & \textbf{0.9526} & \textbf{0.2673} & 0.8834 & 0.6903 & 0.9273 & 0.3283\\
   HoloQA, 2024 \cite{avinab2023hologramQA}  & ImageNet pretrained MoCo-v3 + SHHQ + VGGFace2-HQ & \textbf{0.9144} & \textbf{0.7443} & \textbf{0.9489} & \textbf{0.2671} & \textbf{0.9201} & \textbf{0.7506} & \textbf{0.9523} & \textbf{0.2650}\\
   HoloQA with Frame Tracking, 2024 \cite{avinab2023hologramQA}& ImageNet pretrained MoCo-v3 + SHHQ + VGGFace2-HQ & \underline{\textbf{0.9163}} & \underline{\textbf{0.7497}} & \underline{\textbf{0.9531}} & \underline{\textbf{0.2573}} & \underline{\textbf{0.9291}} & \underline{\textbf{0.7709}} & \underline{\textbf{0.9593}} & \underline{\textbf{0.2497}}\\
   \hline
 \end{tabular}}
\end{table*}

\begin{table*}[!t]
   \renewcommand{\arraystretch}{1.3}
   \caption{Median SRCC, KRCC, PLCC and RMSE of the compared NR video quality models against human judgments of quality of the videos in the LIVE-Meta Rendered Human Avatar VQA database over 100-train-test splits. The \underline{underlined} and \textbf{boldfaced} items represent the best and top three performers, respectively. The deep learning based models contrique, conviqt, and gamival achieved standout performance.}
   \label{tab:NRcomparison}
   \setlength{\tabcolsep}{3pt}
   \centering
   \renewcommand{\tabcolsep}{3.3pt} 
   \resizebox{\textwidth}{!}{
   \begin{tabular}{|c|c|cccc|cccc|}
   \hline
   \multirow{2}{*}{Metrics} & \multirow{2}{*}{Pre-Training / Fine-Tuning Dataset} & \multicolumn{4}{c|}{Body-only} & \multicolumn{4}{c|}{Face-only}\\
   \cline{3-6}\cline{7-10}
   & & SRCC($\uparrow$) & KRCC($\uparrow$) & PLCC($\uparrow$) & RMSE($\downarrow$)& SRCC($\uparrow$) & KRCC($\uparrow$) & PLCC($\uparrow$) & RMSE($\downarrow$) \\
   \hline
   BRISQUE, 2012 \cite{mittal2012no}~ & N/A (\textit{handcrafted}) & 0.8461 & 0.6440 & 0.8934 & 0.3973 &0.8673 & 0.6792 & 0.9135 & 0.3641 \\
   CONTRIQUE, 2022 \cite{9796010} & ImageNet pretrained ResNet-50 + KADIS-700k + AVA + COCO + CERTH-Blur + VOC & \underline{\textbf{0.9092}} & \underline{\textbf{0.7426}} & \underline{\textbf{0.9593}}& \underline{\textbf{0.2511}} & 0.8953 & 0.7102 & 0.9370 & 0.3125 \\
   \hline
   TLVQM, 2019 \cite{korhonen2019two}~  & N/A (\textit{handcrafted})  & 0.5104 & 0.3527 & 0.5966 & 0.7109 & 0.6587 & 0.4719 & 0.7452 & 0.5997\\
   VIDEVAL, 2021 \cite{tu2021ugc}~   & N/A (\textit{handcrafted}) & 0.8497 & 0.6567 & 0.8979 & 0.3914 & 0.8817 & 0.6966 & 0.9220 & 0.3404 \\
   ChipQA, 2021 \cite{9540785} & N/A (\textit{handcrafted}) & 0.8404 & 0.6431 & 0.8853 & 0.4193 & 0.8813 & 0.6967 & 0.9164 & 0.3527 \\
   FAVER, 2022 \cite{zheng2022faver} & N/A (\textit{handcrafted}) & 0.8194 & 0.6277 & 0.8794 & 0.4233 & \textbf{0.9106} & \textbf{0.7391} & \textbf{0.9410} & \textbf{0.3027} \\
   VSFA, 2019 \cite{li2019quality}~    & ImageNet & 0.8670 & 0.6720  & 0.9135  & 0.4079 & 0.8328 & 0.6401 & 0.8904 & 0.4339 \\
   RAPIQUE, 2021 \cite{tu2021rapique}~   & \textit{handcraft} + ImageNet & 0.8469 & 0.6472 & 0.8985 & 0.3933 & \textbf{0.9108} & \textbf{0.7425} & \textbf{0.9505} & \textbf{0.2754} \\
   CONVIQT, 2023 \cite{10243566}~ & ImageNet pretrained ResNet-50 /  Kinetics-400 + Waterloo1k + dareful + REDS + MCML + UVG & \textbf{0.8930} & \textbf{0.7120} & \textbf{0.9532} & \textbf{0.2705} & 0.8834 & 0.6943 & 0.9287 & 0.3285 \\
   GAMIVAL, 2023  \cite{10065464}~   & \textit{handcraft} + ImageNet + GVSET + KUGVD + GISET & ~\textbf{0.8842}~ & ~\textbf{0.6973}~ & ~\textbf{0.9251}~ & ~\textbf{0.3404}~ & \underline{\textbf{0.9203}} & \underline{\textbf{0.7539}} & \underline{\textbf{0.9515}} & \underline{\textbf{0.2719}}\\
   Re-IQA (quality only), 2023 \cite{Saha_2023_CVPR}~   & ImageNet pretrained ResNet-50 \& ImageNet pretrained MoCo-v2 / KADIS-140K + AVA + COCO + CERTH-Blur + VOC  & ~0.8563~ & ~0.6609~ & ~0.9093~ & ~0.3724~ & 0.8381 & 0.6392 & 0.8821 & 0.4161\\
   Re-IQA (content only), 2023 \cite{Saha_2023_CVPR}~   & ImageNet pretrained ResNet-50 \& ImageNet pretrained MoCo-v2 / KADIS-140K + AVA + COCO + CERTH-Blur + VOC & ~0.8476~ & ~0.6536~ & ~0.9070~ & ~0.3755~ & 0.8518 & 0.6586 & 0.9128 &0.3611\\
   Re-IQA (content + quality), 2023 \cite{Saha_2023_CVPR}~   & ImageNet pretrained ResNet-50 \& ImageNet pretrained MoCo-v2 / KADIS-140K + AVA + COCO + CERTH-Blur + VOC & ~0.8715~ & ~0.6806~ & ~0.9271~ & ~0.3327~ & 0.8562 & 0.6620 & 0.9105 & 0.3618\\
   \hline
\end{tabular}}
\end{table*}

\subsection{Evaluation Dataset Processing}

\begin{figure}[!ht]
  \centering
  \footnotesize
  \renewcommand{\tabcolsep}{1.3pt} 
  \renewcommand{\arraystretch}{1.3} 
  \includegraphics[height=150pt]{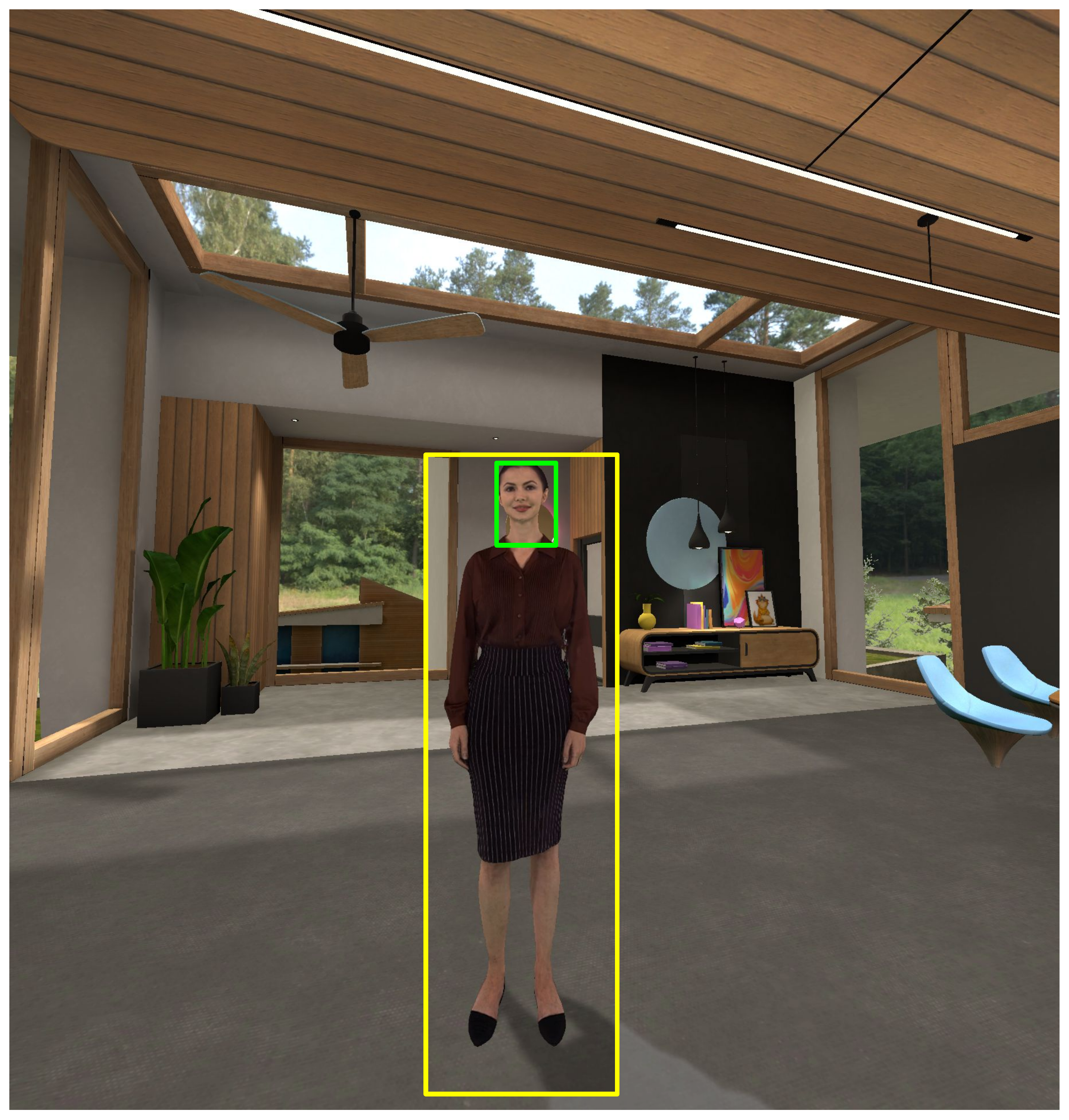}
  \caption{Exemplar of a dumped frame from the human avatar video \textit{Natasha Serious Talking} having original dimension of $1832 \times 1920$ pixels. Within this frame, the yellow bounding box captures the avatar's body ($320 \times 1088$), while the green box surrounds the avatar's face ($94 \times 133$).\label{fig:cropping}} 
\end{figure}

As mentioned in Section \ref{sec:tool}, the Unity tool enabled us to extract frames of both the ground truth and target videos based on user log files. Considering that users may view human avatar videos from various angles, previous research \cite{8743277} has shown that increasing the number of projected views has little correlation to improved quality predictions. Therefore, we adopted a fixed viewing angle when capturing the log files used to generate each frame of the human avatar videos. An example of a dumped frame is depicted in Fig. \ref{fig:cropping}.

Since the background region occupies a large proportion of the displayed content, and may impact the performance of IQA/VQA algorithms, we adopted the method presented in \cite{8743277}, which excludes the background pixels. Likewise, we used the simple expedient of applying the YOLO-v7 model \cite{wang2022yolov7} to extract of bounding boxes around each human avatar. Prior research has demonstrated the significance of frontal views of human bodies and faces as attractors of visual attention \cite{10.1145/1877808.1877819}. To further investigate the impact of facial features and expressions on video quality, we applied the YOLO-v8 face model \cite{Jocher_YOLO_by_Ultralytics_2023} to extract tight bounding boxes around each human avatar's face, within the previously cropped body bounding boxes, as exemplified in Figure \ref{fig:cropping}.

\subsection{Model Evaluation protocol}\label{sec:protocol}

To demonstrate the usefulness of the new LIVE-Meta Rendered Human Avatar VQA Database, we used it to evaluate a variety of leading IQA/VQA algorithms using various standard metrics, including the SRCC, Kendall Rank Correlation Coefficient (KRCC), PLCC, and Root Mean Square Error (RMSE). The SRCC and KRCC measure the degree of monotonicity between the objective model predictions and the human subjective scores, while the PLCC and RMSE gauge the accuracy of the predictions. As usual, a logistic non-linearity function was applied to the predicted quality scores prior to computing the correlations \cite{antkowiak2000final}.

For each split, 80\% of the videos were randomly selected from all the contents to form the training and validation sets, while the remaining 20\% were used as the test set. To maintain fairness of assessment and prevent any model from learning content, we ensured that the subsets did not share any original content.

\subsection{Performance of FR IQA/VQA Models}

In this section, we examine the performance of SOTA FR VQA models on the new LIVE-Meta Rendered Human Avatar VQA Database. As mentioned in Sec. \ref{sec:object-metric}, most existing quality assessment models designed for meshes are FR models, and can be further classified into two types: model-based methods and IQA algorithms that operate on individual frames. The latter are well-suited when only 2D mesh rendering snapshots are to be quality-analyzed. Since there exist few developed FR VQA models designed for 2D mesh projections, we included FR VQA models originally designed to analyze natural videos and for generic VQA tasks.

We comprehensively compared the performance of 17 FR VQA algorithms: PSNR-RGB, PSNR-Y, SSIM \cite{1284395}, DLM \cite{5765502}, VIF \cite{1532311, 1576816}, LPIPS (AlexNet) \cite{Zhang_2018_CVPR}, LPIPS (VGG), CONTRIQUE \cite{9796010}, Re-IQA \cite{Saha_2023_CVPR}, VMAF \cite{li2016toward}, FovVideoVDP \cite{mantiuk2021fovvideovdp}, ST-GREED \cite{Madhusudana_2021}, FUNQUE \cite{9897312}, Y-FUNQUE+ \cite{10375336}, 3C-FUNQUE+ \cite{10375336}, CONVIQT \cite{10243566}, and HoloQA \cite{avinab2023hologramQA} on the new LIVE-Meta database. We calculated the DMOS using Equation \ref{eq:dmos} when computing the performance of the FR VQA models.

Since they do not utilize multiple frames to make VQA predictions at a given moment, PSNR, SSIM, DLM, VIF, LPIPS, CONTRIQUE (FR), and Re-IQA (FR) were calculated on a per-frame basis between the reference videos and corresponding distorted counterparts. These frame-level measurements were subsequently averaged across all frames to obtain aggregate global scores. Among the FR VQA models, PSNR-RGB, PSNR-Y, SSIM, DLM, and VIF are not ordinarily trained and were thus applied directly on all 1000 test sets. We used the pre-trained open-source version of VMAF (v0.6.1) originally designed for general-purpose VQA tasks. Likewise, we report the results obtained with the publicly available calibrated FovVideoVDP model. When implementing LPIPS, CONTRIQUE (FR), Re-IQA (FR), VMAF, ST-GREED, FUNQUE, Y-FUNQUE+, 3C-FUNQUE+, CONVIQT (FR), and HoloQA, features were extracted and an SVR was trained using 80\%/20\% train/test sets. CONTRIQUE (FR), ReIQA (FR), and CONVIQT (FR) are full-reference implementations of established unsupervised NR models, while HoloQA is a new hybrid model that combines both traditional and deep learning-based features for comprehensive quality assessment. The optimal parameters of the SVR were determined using a five-fold cross-validation procedure on the training and validation sets.

\subsection{Performance of NR IQA/VQA Models}

Table \ref{tab:FRcomparison} provides an overview of the median performance of the above FR IQA/VQA algorithms on the LIVE-Meta Rendered Human Avatar VQA Database. The standout performers were CONTRIQUE (FR), ReIQA (FR), and CONVIQT (FR), which generalize well since they are unsupervised, and the two versions of HoloQA, which were designed to analyze human avatar content.

We also conducted a comprehensive evaluation to gauge the performances of existing NR IQA/VQA algorithms on the new LIVE-Meta database. A selection of prominent generic NR IQA/VQA models, namely BRISQUE \cite{mittal2012no}, CONTRIQUE, TLVQM \cite{korhonen2019two}, VIDEVAL \cite{tu2021ugc}, ChipQA \cite{9540785}, FAVER \cite{zheng2022faver}, VSFA \cite{li2019quality}, RAPIQUE \cite{tu2021rapique}, CONVIQT, GAMIVAL \cite{10065464}, and Re-IQA (quality only, content only, content + quality) \cite{Saha_2023_CVPR},  were tested. BRISQUE, TLVQM, VIDEVAL, ChipQA, and FAVER are primary based on quality-aware neurostatistic distortion models, while RAPIQUE and GAMIVAL fuse neurostatistical features with deep learning-based features. CONTRIQUE derives from a self-supervised learning framework that aims to learn quality-aware representations without the need to train on human opinion scores. The VSFA model leverages deep learning to extract features, which are subsequently mapped to MLE-MOS. CONVIQT combines spatial CONTRIQUE features with temporal quality features, also without supervision. Re-IQA employs a multi-modal approach that integrates quality and content-aware features derived from pre-trained deep learning models to enhance prediction accuracy.

The extraction of quality-aware features in frame-based models like BRISQUE and CONTRIQUE was performed on a per-frame basis, which were then pooled over frames to obtain quality predictions. The supervised methods, including BRISQUE, CONTRIQUE, TLVQM, VIDEVAL, ChipQA, FAVER, RAPIQUE, CONVIQT, GAMIVAL, and Re-IQA all employed an SVR to map the pooled and combined quality-aware features to MLE-MOS. GAMIVAL, which was designed for conducting VQA on gaming videos, modifies RAPIQUE by employing deep pre-trained gaming content model called NDNet-Gaming \cite{utke2020ndnetgaming}. We followed the same evaluation protocol using 80\%/20\% train/test splits.

Table \ref{tab:NRcomparison} summarizes the median performances of the compared NR IQA/VQA algorithms on the new VQA database. TLVQM, which relies on multiple hand-tuned hyperparameters optimized for predicting generic video quality, was unable to generalize to human avatar videos.

However, algorithms leveraging neurostatistical video distortion features, including BRISQUE, VIDEVAL, ChipQA, FAVER, RAPIQUE, and GAMIVAL, all performed well. While VIDEVAL slightly outperformed RAPIQUE on the body-only evaluation, RAPIQUE demonstrated superior performance on the face-only evaluation. The resizing of frames in the body-only evaluation could lead to performance degradations in RAPIQUE, which also uses aggressive frame sub-sampling. Incorporating deep learning techniques in models such as CONTRIQUE, VSFA, RAPIQUE, CONVIQT, GAMIVAL, and Re-IQA yielded substantial performance improvements. This underscores the ability of learning-based approaches to capture the inherent statistical structure of synthetically generated human avatar videos and their distortions.

\section{Conclusion and Discussion} \label{conclusion}

We have presented the LIVE-Meta Rendered Human Avatar VQA database, which is a new resource for the development and evaluation of FR and NR VQA algorithms that are specifically designed for VR textured mesh content. Although we cannot make the proprietary Metastage videos freely available, other users may also purchase them. To facilitate such efforts, we also makes the metadata of the database publicly available at \href{https://live.ece.utexas.edu/research/LIVE-Meta-rendered-human-avatar/index.html}{https://live.ece.utexas.edu/research/LIVE-Meta-rendered-human-avatar/index.html}.

Our study utilized Oculus Quest Pro headsets, which may limit the generalizability of our findings to other VR environments. Future research would benefit by exploring and comparing diverse VR environments and headsets to obtain a broader understanding of the device factors affecting VR video quality.

We showed the usefulness of the new database for analyzing, benchmarking, and designing for FR and NR IQA/VQA algorithms.  Future investigations may concentrate on the development of deep learning methodologies to enhance the performance of FR and NR IQA/VQA algorithms for avatar analysis.

While our dataset includes color resolution, depth resolution, delay artifacts, and frame rate distortions, other distortion types can be generated using our tool, such as color errors, edge noise, hole frequency, and surface noise. Future studies that include these additional distortion types would be useful.

\section*{Acknowledgments}

The authors extend their gratitude to the human study participants for their valuable contributions and time in the human study. They also thank The University of Texas at Austin for their support and resources. Special appreciation is expressed to Metastage for providing the dataset, which was essential for conducting the study and obtaining valuable insights. The authors are grateful for Metastage's contribution and support in advancing the field of VR human avatar video research.


\bibliographystyle{IEEEtran}
\bibliography{IEEEexample}{}

\end{document}